\documentclass[12pt]{article}
\usepackage{amsmath,amssymb,amsfonts,amsthm,graphicx,placeins, float, hyperref}
\usepackage{color}
\usepackage{multirow,booktabs}

\title{Flexible Bayesian modeling of counts: constructing penalized complexity priors}

\newcommand{\y}{\mbox{\boldmath$y$\unboldmath}}
\newcommand{\bbeta}{\mbox{\boldmath$\beta$\unboldmath}}
\newcommand{\btau}{\mbox{\boldmath$\tau$\unboldmath}}
\newcommand{\bet}{\mbox{\boldmath$\eta$\unboldmath}}

\newcommand{\x}{\mbox{\boldmath$x$\unboldmath}}

\newcommand{\Z}{\mbox{\boldmath$Z$\unboldmath}}

\newcommand{\bK}{\mbox{\boldmath$K$\unboldmath}}

\newcommand{\btheta}{\mbox{\boldmath$\theta$\unboldmath}}

\newcommand{\zero}{\mbox{\boldmath$0$\unboldmath}}

\newcommand{\f}{\mbox{\boldmath$f$\unboldmath}}

\newcommand{\bpsi}{\mbox{\boldmath$\psi$\unboldmath}}

\newcommand{\RNum}[1]{\uppercase\expandafter{\romannumeral #1\relax}}

\newtheorem{Theorem}{Theorem}[section] 

\newtheorem{thm}[Theorem]{Theorem}

\newtheorem{rem}[Theorem]{Remark}

\usepackage{amsmath}
\usepackage{stackengine}
\begin{document}
\maketitle

\author{Mahsa Nadifar$^{*,\ddagger}$, Hossein Baghishani$^*$, Thomas Kneib$^{\ddagger}$, Afshin Fallah$^{**}$\\ 
$^*$ Department of Statistics, Faculty of Mathematical Sciences, Shahrood University of Technology, Iran\\
$^{\ddagger}$ Georg-August-University G\"{o}ttingen, Germany\\
 $^{**}$ Department of Statistics, Imam Khomeini International University, Qazvin, Iran}

\section*{Abstract}
Many of data, particularly in medicine and disease mapping, are count. Indeed, under- or over-dispersion problem in count data distrusts the performance of the classical Poisson model. For taking into account this problem, in this paper, we introduce a new Bayesian structured additive regression model, called gamma count, with enough flexibility in modelling dispersion. Setting convenient prior distributions on the model parameters is a momentous issue in Bayesian statistics that characterize the nature of our uncertainty parameters. Relying on a recently  proposed class of penalized complexity priors, motivated from a general set of construction principles, we derive the prior structure. The model can be formulated as a latent Gaussian model, and consequently, we can carry out the fast computation by using the integrated nested Laplace approximation method. We investigate the proposed methodology simulation study. Different expropriate prior distribution are examined in order to provide a reasonable sensitivity analysis. To explain the applicability of the proposed model, we analyzed two real word data sets related to larynx mortality cancer in Germany and handball champions league.  \\
\noindent{\bf Keywords:} 
\begin{small}
Structured additive regression model, over-dispersion, Penalized Complexity prior,  count, under-dispersion, Gamma Count, INLA.\\
\end{small}
\noindent{\bf Mathematics Subject Classification (2010):} $\rm 62-06$, $\rm 62J12$, $\rm 62F15$,  $\rm 62H11$.\\

\section{Introduction}
The Poisson regression is the prevalent model for analyzing count data, especially when provided by a structured additive regression (STAR; Fahrmeir et al., 2004; Kneib et al., 2009). STAR modeling allows to simultaneously incorporate model components for non-linear, temporal, and spatial effects. Although the Poisson regression model is widespread, it limits the conditional variance to be equal to the conditional mean, which leads to an equi-dispersion situation. Such a case is seldom observed in many practical data analyses. Indeed, in real-life applications, count data can exhibit different aspects, specifically under-dispersion (the mean exceeding the variance) and over-dispersion (the variance exceeding the mean).

Over the years, several extended count regression models have been developed for dealing with non-equivalent dispersion. Some approaches, such as adopting a generalized linear mixed model (GLMM; Breslow and Clayton, 1993), embedded a random effect in the model to consider the over-dispersion of counts. A convenient proposal, in this direction, is a Poisson model with gamma-distributed random effects yielding to a negative binomial (NB) model. However, the NB model is not a good candidate for analyzing under-dispersed counts (Cameron and Trivedi, 2013). Two alternative classes of models for accounting unobserved heterogeneity are finite mixture models (Pearson, 1894) and hurdle models (see, for example, Cameron and Trivedi, 2013; Baetschmann and Winkelmann, 2017). Indeed, we can use the hurdle model for accounting both over-dispersion and under-dispersion in the counts. Other methods incorporate weighting the Poisson distribution (Ridout and Besbeas, 2004), the COM-Poisson distribution (Lord \textit{et al.}, 2010; Lord \textit{et al.}, 2008), and the generalized Poisson inverse Gaussian family (Zhu and Joe, 2009) to name a few. Generally, analysis of under-dispersed counts has received less attention (Zeviani \textit{et al}., 2014).

Winkelmann (1995) proposed a new methodology for over- or under-dispersed counts based on the renewal theory (Cox, 1962) that relates non-exponential durations (inter-arrival/waiting times) between events and the distribution of the counts. He connected the models for counts and models for durations relaxing the assumption of equi-dispersion at the cost of an extra parameter. Winkelmann (1995) replaced independently and identically exponentially distributed waiting times (which would lead to the Poisson distribution for counts) by a less restrictive non-negative distribution with a non-constant hazard function. The hazard function is the crucial quantity when studying renewal processes; it completely characterizes the distribution of inter-arrival times and describes the type of dispersion observed in the corresponding count data. Winkelmann (1995) showed that if the hazard function is monotonic, increasing (decreasing) hazard corresponds to count data with under-dispersion (over-dispersion). Therefore, providing a more flexible hazard function results in a more flexible counting process able to support over-dispersed and under-dispersed, as well as equi-dispersed data. Several researchers have introduced some models considering this methodology, including a gamma-count (GC) model (Winkelmann, 1995; Toft \textit{et al.}, 2006; Zeviani \textit{et al.}, 2014), Weibull-count model (McShane \textit{et al.}, 2008), and lognormal-count model (Gonzales-Barron and Butler, 2011). Nadifar \textit{et al.} (2019) extended the GC model to analyze spatially correlated count data in a Bayesian framework using integrated nested Laplace approximation (INLA; Rue \textit{et al.}, 2009). They used the GC spatial regression model for the analysis of groundwater quality data.

In this article, we aim to further generalize the suggested model in Nadifar et al. (2019) to a gamma-count structured additive (GCSA) regression model by developing a new class of penalized complexity (PC) prior distributions, introduced by Simpson et al. (2017), for the dispersion parameter as well as precision parameters of Gaussian priors considered for various effects in the predictor. PC priors are informative and have desirable aspects, such as invariance under reparameterizations, connection to Jeffreys' prior, and enjoying robustness properties. As Simpson et al. (2017) have noticed, their proposed approach uses the natural nested structure of many model components, which represents the model component as a flexible extension of a base model. Therefore, deviations from the base model increase the complexity of the model. For example, the t-distribution model can be regarded as a generalization of the normal distribution, as
the base model, for the limiting case of increasing degrees of freedom. An interesting base model is obtained in STAR models when, e.g. for a non-linear effect, setting the smoothing variance to zero.

In short, Simpson \textit{et al.} (2017) developed the PC prior by the following steps: 1) The increased complexity between the more flexible model and the base model is measured using the Kullback-Leibler divergence, 2) The deviation from the base model is penalized with a constant decay rate. The base model would be favored until there is enough support for a more complex model. Therefore, an exponential prior is assigned to the Kullback–Leibler distance such that the mode of the prior corresponds to the base model. 3) The decay rate is determined by controlling the prior mass in the tail. In STAR models, the rate can be extracted based on prior assumptions about the scaling of the model components. That is why Klein and Kneib (2016) referred to the priors as scale-dependent priors. 
They extended scale-dependent priors for the variance parameters in structured additive distributional regression models and compared them by some accepted alternative prior distributions. Klein and Kneib (2016) developed Markov chain Monte Carlo simulation inference with scale-dependent priors by constructing proposal densities based on the idea of iteratively weighted least squares (Gamerman, 1997; Brezger and Lang, 2006).

The proposed inferential framework of Klein and Kneib (2016) could be comprehensive, as it incorporates scale-dependent priors not only in mean regression models for responses from the exponential family but rather considers the general framework of distributional regression (Klein, Kneib and Lang, 2015; Klein, Kneib, Lang and Sohn, 2015) where further moments or general shape parameters of the conditional response distribution can be linked to a predictor. Despite these advantages, we prefer to use the integrated nested Laplace approximations (INLA) method introduced by Rue \textit{et al.} (2009) due to the wide range of problems of MCMC algorithms regarding convergence and computational time when applied to complex models like our proposed model. To overcome the difficulties associated with the MCMC methods, Rue \textit{et al.} (2009) introduced the INLA, a very fast, non-sampling-based approximate Bayesian methodology. This method combines Laplace approximations and numerical integration efficiently for a particular class of models, the so-called latent Gaussian models. INLA substitutes MCMC simulations with accurate, deterministic approximations to posterior marginal distributions. The R package \texttt{R-INLA} is free for download from \url{http://www.r-inla.org/} and fairly easy to use. Many examples of applications in several fields have also appeared in the recent literature (Bakka \textit{et al.}, 2018; Rue \textit{et al.}, 2017; S{\o}rbye \textit{et al.}, 2018; Simpson \textit{et al.}, 2016; Muff \textit{et al.}, 2015; Blangiardoa \textit{et al.}, 2013; Martins \textit{et al.}, 2013; Schr\"{o}dle and Held, 2011; Paul \textit{et al.}, 2010; Martino and Rue, 2010). 

The plan for the rest of this paper is as follows. In Section \ref{Sec2}, we demonstrate the GCSA regression model and the derivation of the new prior structures. The fundamental methodology for Bayesian analysis of GCSA regression is described in Section \ref{Sec4}. The performance of the proposed approach is examined in simulation studies under various scenarios. The basic results are briefly summarized in Section \ref{Sec5}. In contrast, all simulations are documented in more detail in the supplementary materials. Section \ref{Sec6} applies the methodology to two real datasets: Larynx cancer mortality counts in Germany, as an under-dispersed data example; and handball match data, as an over-dispersed data example. Finally, we discussed the results in section \ref{Sec7}.

\section{\label{Sec2} Penalized Complexity Priors for the GCSA Model}
Our goal is to construct a PC prior for the dispersion parameter and scale-dependent hyperprios for the precision parameters of the Gaussian priors considered for various effects in the structured additive predictor of a GCSA regression model. In the proposed model, the dispersion is treated as a fixed parameter.

\subsection{Observational Model}
In this section, we provide a brief background on the definition and properties of the GC distribution  and its underlying classical regression model for uncorrelated data. 

As Winkelmann (2008) has noticed, the count and the duration view are just two different representations of the same underlying stochastic process. From a statistical viewpoint, the distribution of cumulative waiting times uniquely determines the distribution of counts and vice versa. This relationship can be employed to derive new count data distributions (Winkelmann, 1995; McShane \textit{et al.}, 2008, Gonzales-Barron and Butler, 2011; Ong \textit{et al.}, 2015). For example, the Poisson distribution corresponds to the exponential inter-arrival times between events. The GC distribution has been proposed based on gamma distributed inter-arrival times by Winkelmann (1995).

Let $u_k$ be the waiting time between the $(k-1)$th and $k$th events. Therefore, the arrival time of the $n$th event is given by
\[S_n=\sum_{k=1}^n u_k,~~~n=1,2,\cdots .\]
Let $Y(t)$ denote the total number of events that have occurred in the interval $(0,t)$. Hence, $\{Y(t), ~ t>0\}$ is a counting process and for a fixed $t$, $Y(t)$ is a count variable. The stochastic properties of the counting process (and consequently of the count variable) are entirely determined once we know the joint distribution function of the waiting times, $\{u_k,~k\geq 1\}$. In particular, $Y(t)<n$ if and only if $S_n>t$. Therefore,
\[{\rm P}(Y(t)<n)={\rm P}(S_n>t)=1-F_n(t),\]
where $F_n(t)$ is the comulative distribution function of $S_n$. Moreover, 
\[{\rm P}(Y(t)=n)=F_n(t)-F_{n+1}(t).\]
Generally, $F_n(t)$ is a complicated convolution of the underlying densities of $u_k$'s, which makes it analytically intractable. However, by using the theory of renewal processes, a significant simplification arises if $u_k$'s are independently and identically distributed with a standard distribution.

We assume that $\{u_k,~k\geq 1\}$ is a sequence of independently and identically gamma distributed variables, $Gamma(\alpha, \gamma)$, with mean ${\rm E}(u_k)=\alpha/\gamma$ and variance ${\rm Var}(u_k)=\alpha/\gamma^2$.
It can be shown that if $Y(t)$ denotes the number of events within $(0,t)$ interval, it follows a GC distribution with parameters $\alpha$ and $\gamma$, denoted by $Y(t)\sim GC(\alpha, \gamma)$. Here, $\alpha$ is the dispersion parameter of distribution and controls the amount of dispersion in the counts. 
The probability mass function of $Y(t)$ is given by 
\begin{eqnarray}\label{f2}
{\rm P}(Y(t)=y)=G(y\alpha,\gamma t)-G((y+1)\alpha,\gamma t), ~~~~~ y=0,1,2,\cdots,
\end{eqnarray}
where
\[G(n\alpha,\gamma t)=\frac{1}{\Gamma(n\alpha)}\int_0^{\gamma t}x^{n\alpha-1}e^{-x}dx,~~~~~n=1,2,\cdots.\]
For non-integer $\alpha$, no closed form expression is available for $G(y\alpha,\gamma t)$ and thus for ${\rm P}(Y(t)=y)$.
Also, for $\alpha=1$, the distribution of $u_k$ reduces to the exponential, and \eqref{f2} simplifies to the Poisson distribution. Therefore, the GC distribution could be considered as a flexible extension of the Poisson distribution, as a base model, when setting the dispersion parameter to one. 
More importantly, when $0<\alpha<1$ ($\alpha>1$) the GC distribution is over-dispersed (under-dispersed).
Moreover, the mean of GC distribution is given by
\begin{eqnarray}\label{f3}
{\rm E}(Y(t))=\sum_{k=1}^{\infty}G(k\alpha ,\gamma t).
\end{eqnarray}
We refer readers to Winkelmann (1995) for more details about properties of the GC distribution. 

\subsubsection{Regression Model based on GC Distribution}
We assume that GC distribution for count observations $y_i$ as well as covariates information $\x_i=(1, x_{i1},\ldots,x_{ip})^\prime$ have been collected for individuals $i=1,\ldots , n$.  
The mean of GC distribution in \eqref{f3} has no closed form. Therefore, extending a regression model based on the mean is not straightforward. Assuming that the length of the time interval is the same for all observations, we can set $t$ to unity, without loss of generality.
This results in the following regression model (Zeviani et al., 2014): 
\begin{eqnarray}\label{f4} 
\log\left({\rm E}\left(u_{k_i}|\x_i\right)\right)&=&\log\left(\frac{\alpha}{\gamma\left(\x_i\right)}\right)\cr 
&=&-(\beta_0+x_1\beta_1+\ldots +x_p\beta_p) = - \eta_i, ~~~~~~i=1,\ldots , n, 
\end{eqnarray} 
where $\bbeta=(\beta_0,\beta_1,\ldots,\beta_p)^\prime\in \mathbb{R}^{p+1}$ is the vector of regression coefficients. 
We should notice that the regression model is defined on the waiting times $u_{k_i}$ instead of $Y_i$, where $u_{k_i}$ is the generic representation of waiting times for the $i$th observation. Its origin is for failure to establish the equality ${\rm E}(Y_i|\x_i)=\big({\rm E}(u_{k_i}|\x_i)\big)^{-1}$ unless for $\alpha=1$. Indeed, given the inverse relationship between gaps and the number of occurrences, 
the minus sign behind $\bbeta$ is due to the reverse effect of covariates on waiting times instead of counts; the longer the expectation of time interval, the fewer the number of occurrences.  
Since $\{u_{k_i},~k\geq 1\}$ are independently and identically distributed random variables, we drop the index ${\rm k}$ without loss of generality. 
For given covariates $\x_i$, the responses $y_i$ are conditionally independent where their conditional mean ${\rm E}(Y_i|\x_i)$ can be evaluated via equation \eqref{f3}. 
From \eqref{f4}, one can write 
\begin{eqnarray*} 
\gamma(\x_i) =\alpha\exp(\eta_i),~~~~~~~i=1,\ldots ,n. 
\end{eqnarray*} 
Therefore, given a sample of independent observations $\{(y_i,\x_i),i=1,\ldots, n\}$, the GC regression model can be written as 
$y_i|\alpha,\eta_i\sim GC(\alpha,\alpha\exp(\eta_i))$, $i=1,\ldots ,n$.
The likelihood function is given by 
\begin{eqnarray*} 
{\rm L}(\alpha,\bet|\y)&=&\prod_{i=1}^{n} \{G(\alpha y_{i},\alpha\exp(\eta_i))-G(\alpha y_{i}+\alpha ,\alpha\exp(\eta_i))\}, 
\end{eqnarray*} 
where 
$\y=(y_1,\ldots,y_n)^\prime$ and $\bet=(\eta_1,\ldots,\eta_n)^\prime$ are the vector of observed counts and linear predictor, respectively. 
There are no explicit forms for the maximum likelihood estimators of the parameters, due to the complexity of the likelihood function. Therefore, numerical optimization is needed for estimating the model parameters (Winkelmann, 1995; Zeviani \textit{et al.}, 2014). 
\subsection{The Proposed Model}
For constructing the GC structured additive regression model, we assume that the conditional distribution of response variable $y_i$, $i=1,\ldots ,n$,  has density
\begin{eqnarray}\label{f5}
y_i| \alpha,\eta_i\sim GC\left(\alpha, \alpha\exp\left(\eta_i\right)\right),~~~~~~i=1,\ldots,n,
\end{eqnarray}
where $\eta_i$ is the structured additive predictor as 
\begin{eqnarray}\label{ff6}
\eta_i=\beta_{0}+\sum_{j=1}^{J}f_{j}(\x_i)
\end{eqnarray}
in which $\beta_0$ is an intercept term representing the overall level of the predictor. The functions $\{f_j\}$ show different functional effects, including the time effect, the spatial effect, and the non-linear effect of a continuous covariate. Each function $f_{j}(\x)$ depending on (different subsets of) $\x$, is represented by a linear combination of basis functions. Ignoring the index $j$, a typical function $f(\x)$ is then specified as
\[f(\x)=\sum_{d=1}^D\beta_dB_d(\x),\]
where $B_d(\x)$, $d=1,\ldots,D$, is a set of appropriate basis functions and $\bbeta=(\beta_1,\ldots,\beta_D)^\prime$ is the vector of corresponding basis coefficients. Indeed, to ensure identifiability, the functional effects are centered about zero.
Using this approach, we can express all vectors $\f_{j} = (f_{j}(\x_1),\ldots, f_{j}(\x_n))^\prime$, $j=1,\ldots,J$, as the matrix product of an appropriately defined design matrix $\Z_{j}$ and a vector $\bbeta_j$ of unknown parameters, i.e. 
$\f_{j} = \Z_{j}\bbeta_{j}$. See Fahrmier \textit{et al}. (2013) and Klein and Kneib (2016) for more details.

In a Bayesian model specification, priors for $\f_j$ are specified through appropriate design matrices and priors for the random effects $\bbeta_j$. See Fahrmeir and Echavarria (2006) for details on defining the appropriate design matrices for crucial predictor components, including P-splines and Markov random fields. Priors for $\bbeta_j$ are (partially improper) multivariate Gaussian
\begin{eqnarray} 
\label{fprec} 
p(\bbeta_j|\sigma^2_j)\propto (\sigma^2_j)^{-\frac{rk(\bK_j)}{2}}\exp\left(\frac{-1}{2\sigma^2_j}\bbeta_j^{'} \bK_j \bbeta_j\right),~~~~j=1,\ldots, J, 
\end{eqnarray} 
with fixed positive (semi-)definite precision matrix $\bK_j$, and $rk(\cdot)$ denoting the rank of matrix. The matrix $\bK_j$ is chosen to penalize roughness of the function. Its structure depends on the type of covariate and the smoothness of the function. The variance parameter $\sigma^2_j$ quantifies our prior uncertainty about the properties enforced by $\bK_j$. It corresponds to a smoothing parameter in a frequentist setting (Fahrmeir and Echavarria, 2006). For intercept $\beta_0$, we typically assume a zero-mean Gaussian prior with the precision equal to $0.01$. 

For completing the model specification, a suitable prior for dispersion parameter, $\alpha$, and hyperpriors for the variance parameters, $\sigma^2_j$, have to be provided. Constructing a suitable prior for $\alpha$ has been less noticed until now. Nadifar \textit{et al.} (2019) considered the $Gamma(1,0.005)$ prior and it was fixed as prior assumptions in the \texttt{R-INLA} package. Since the Poisson distribution is a nested model of the GC distribution, in this paper, we apply the approach of Simpson et al. (2017) to define a PC prior for $\alpha$ as well. To this end, we employ the distributional structure of inter-arrival times. We compare the proposed PC prior with the gamma prior as well.

Given a comprehensive debate on the appropriateness of the gamma distribution as a natural conjugate hyperprior for $\sigma^2_j$, Klein and Kneib (2016) developed scale-dependent priors by extending the principled approach of Simpson et al. (2017) in structured additive distributional regression models. We will use their results to specify the scale-dependent hyperpriors given the dispersion parameter of the GCSA model. We also consider generalised beta prime, and inverse gamma priors for the variances that have been suggested as hyperpriors in Bayesian mixed-effects models (Gelman, 2005, 2006; Hodges, 2013; Klein and Kneib, 2016) and compare their performances. One of the advantages of our proposed model is ensuring that the posterior would be proper.  It would be essential because priors for the vectors of regression coefficients for several effect types, such as Markov random fields, in the structured additive regression are partially improper.

\subsection{Derivation of the Prior Structure}
We will now explain how to construct the PC prior for dispersion parameter and scale-dependent hyperpriors for the variance parameters. Let $\xi$ is the parameter of interest we want to define a prior. According to Simpson et al. (2017), principles for constructing a PC prior distribution are as follows: 
\begin{itemize} 
\item[~] \textbf{Principle 1:} 
\textit{Occam’s razor.} Based on the principle of parsimony, to build a prior, a simple base model for each effect is supported unless the data give enough evidence for a more complex modeling alternative. \\ 
\item[~]\textbf{Principle 2:} \textit{Measure of complexity.} The increased complexity between the base model represented by density $p_b$ and the alternative complex model represented by density $p$ is measured by the unidirectional measure 
\[d := d(p ||p_b)=\sqrt{2{\rm KLD}(p ||p_b)}\] 
where ${\rm KLD}(p||p_b)$ is the Kullback–Leibler divergence given by
\begin{eqnarray*} 
{\rm KLD}(p||p_b)=\int p(x)\log\left(\frac{p(x)}{p_b(x)}\right)dx. 
\end{eqnarray*} 
\item[~]\textbf{Principle 3:} \textit{Constant rate penalization.} This principle implies an exponential prior 
$p(d) = \lambda \exp(-\lambda d)$, with a constant decay-rate $r=\exp(-\lambda)$, on the distance scale $d$ obtained from Principle 2. The mode of the prior is at $d = 0$, i.e. the base model is in favor. Using the change of variable theorem, the prior on the original space is as follows:
\[p(\xi)=\lambda e^{-\lambda d(\xi)}|\frac{\partial d(\xi)}{\partial\xi}|.\]
\item[~]\textbf{Principle 4:} \textit{User-defined scaling.} Using this principle, we can select $\lambda$ by controlling the prior mass in the tail. This condition is of the form
\begin{eqnarray} 
\label{P4} 
{\rm P}\left(q\left(\xi\right)\leq u\right) = 1-a 
\end{eqnarray} 
where $q(\cdot)$ is an interpretable transformation of $\xi$ and $u > 0$ and $a \in (0, 1)$ are some user-defined values. The probability in \eqref{P4} depends on the intensity $\lambda$ through the density of $q(\xi)$ such that solving the expression for $\lambda$ yields the exact prior specification for $\xi$. It also allows the user to prescribe how informative the resulting PC prior is. We will discuss about the choice of $q(\cdot)$, $u$ and $a$ in the following.
\end{itemize} 
A significant advantage of PC priors is the reparameterization-invariance property since the prior is defined on the distance $d$, which is then transformed to the corresponding prior for $\xi$. We first define the PC prior for $\alpha$. Let $GC(\alpha,\beta)$, be the GC distribution with the dispersion parameter $\alpha$. When $\alpha=1$, the GC distribution reduces to the Poisson, the simple base model here. Considering dual modeling of counting and timing processes, we define flexible and base models based on the inter-arrival distribution.
\begin{thm}\label{thm1} 
Let $Gamma(\alpha, \beta_1)$ be the flexible model, and $Gamma(1, \beta_2)$ is the base model.The PC prior for $\alpha$ is defined as follows 
\begin{eqnarray} 
\label{pcalpha} 
p(\alpha)&=&\lambda e^{-\lambda \sqrt{-2\log\Gamma(\alpha)+2(\alpha-1)\psi(\alpha)+2\log\frac{\beta_1}{\beta_2}-2\alpha(1-\frac{\beta_2}{\beta_1})}}\cr 
&~~&|\frac{\left(\alpha-1\right)\psi'\left(\alpha\right)-(1-\frac{\beta_2}{\beta_1})}{\sqrt{-2\log\Gamma(\alpha)+2(\alpha-1)\psi(\alpha)+2\log\frac{\beta_1}{\beta_2}-2\alpha(1-\frac{\beta_2}{\beta_1})}}|. 
\end{eqnarray} 
\end{thm} 
To prove the theorem, we have to adjust the Kullback-Leibler divergence between two gamma distributions, which is simple to measure (Abdulrahman et al., 2015). The details are provided in the supplementary file.
\begin{rem}\label{rm1} 
According to the definition of GC distribution, provided $\f_j$s are given, the rate parameters of base and flexible models should be equal, i.e. $\beta_1=\beta_2=\beta$. Therefore, the proposed PC prior density for $\alpha$ reduces to a simplified form as follows 
\begin{eqnarray} 
\label{PCalpha2} 
p(\alpha)&=&\lambda e^{-\lambda \sqrt{-2\log\Gamma(\alpha)+2(\alpha-1)\psi(\alpha)}}\cr 
&~~&|\frac{\left(\alpha-1\right)\psi'\left(\alpha\right)}{\sqrt{-2\log\Gamma(\alpha)+2(\alpha-1)\psi(\alpha)}}|. 
\end{eqnarray} 
\end{rem} 
\begin{rem}\label{rm1-2} 
\textbf{User-defined scaling.} The shrinkage parameter, $\lambda$, has to be determined by solving 
${\rm P}(q(\alpha) > u) = a$. By considering $q(\alpha) = d(\alpha)$, we have 
\begin{eqnarray} 
\label{P4GC} 
{\rm P}(q(\alpha) < u) &=&\int_{0}^{u} \lambda e^{-\lambda \sqrt{-2\log\Gamma(\alpha)+2(\alpha-1)\psi(\alpha)}} d\alpha\cr 
&=& 1-a.
\end{eqnarray} 
\end{rem} 
Equation \eqref{P4GC} could be easily solved using numerical integration. We can also determine hyperparameters based on our beliefs (Simpson et al., 2017). Considering the main feature of the GC distribution, i.e. modeling under- (over-)dispersed data, we can select some optimal values for the hyperparameters based on a simulation study. We describe it in Section 3.2. The scale-dependent hyperprior for the generic variance parameter $\sigma^2$ is given in the following theorem.
\begin{thm}\label{thm2} 
Let $\bbeta\sim \mathtt{N}_{rn(\bK)}(\zero, \frac{1}{\tau}\bK^{-1})$. Assume $\frac{1}{\sigma^2}\bK$ be the precision matrix of the flexible model for a vector of 
parameters $\bbeta$ and $\frac{1}{\sigma^2_b}\bK$ the precision matrix of the base model where $\sigma^2_b \rightarrow 0$. 
Furthermore, let $p(\sigma^2)$ be the prior for $\sigma^2$ depending on a hyperparameter $\epsilon$. Then 
\begin{eqnarray} 
\label{pctau} 
p(\sigma^2)=\frac{1}{2\epsilon}\left(\frac{\sigma^2}{\epsilon}\right)^{-1/2}\exp\left(-\left(\frac{\sigma^2}{\epsilon} \right)^{1/2}\right),
\end{eqnarray} 
which is a Weibull prior with shape parameter $a = 1/2$ and scale parameter $\epsilon$.
\end{thm} 
Proof of this theorem is conveniently taken from the proof in Appendix A2 by Simpson et al. (2017) by changing variable. Klein and Kneib (2016) provide the details of proof in Supplement A.1.
 Following Principle 4, we can infer $\epsilon$ from a notion of scale by considering ${\rm P}(\sigma^2 > u) =a$. It would assume that large standard deviations are less likely. Consequently, we have $\epsilon=-\ln(a)/u$ (Simpson \textit{et al.}, 2017). 

As indicated by Simpson et al. (2017), the invariance property of the prior is guaranteed. For example, 
type-2 Gumbel distribution is obtained as the prior for the precision parameter $\xi = 1/\sigma^2$.

\subsection{Competing Priors}\label{Sec2.3}
\begin{figure}[!htbp] 
\begin{center} 
\begin{tabular}{cc}  
\includegraphics[width=8cm,height=8cm,keepaspectratio]{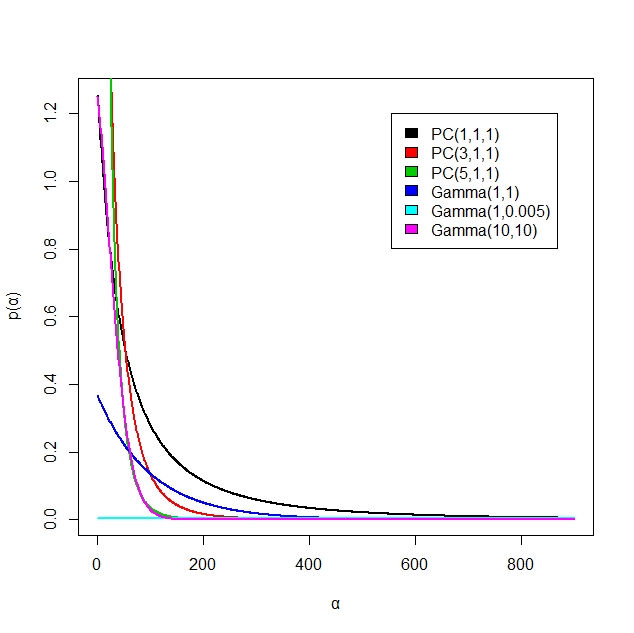} & \includegraphics[width=8cm,height=8cm,keepaspectratio]{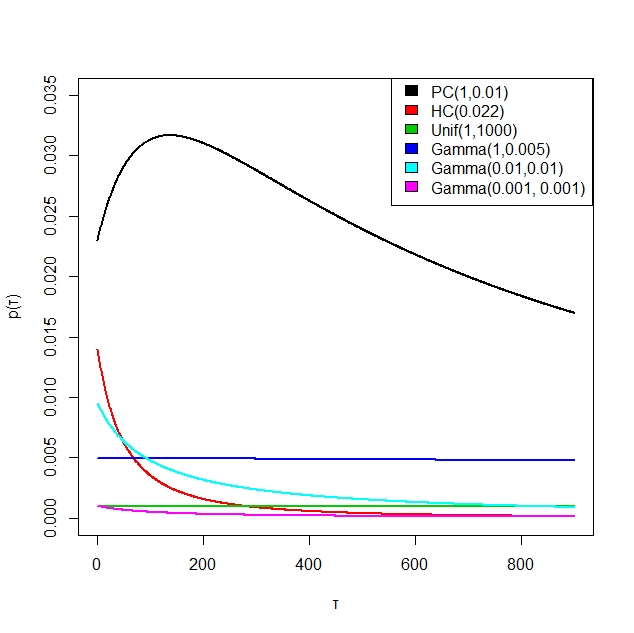}\\ 
(a)& (b)\\
\end{tabular} 
\caption{\label{figp} Illustration of densities for different priors and hyperpriors for $\alpha$ and $\tau$. The panel (a) shows the priors $\pi(\alpha)$, the panel (b)  shows the hyperpriors $\tau$.} 
\end{center} 
\end{figure}

To analyze the sensitivity of the various priors of parameters, 
we consider some alternative priors following previous researches and based on simulation studies. A brief explanation of the parameters are in Table \ref{Tabprior}: 
\begin{table}[H] 
\centering \caption{\label{Tabprior} Overview of available hyperpriors for $\alpha$ and $\sigma^2$} 
\begin{tabular}{l|ll} 
\toprule 
\textbf{Name}&\textbf{Prior}&\textbf{Information}\\ 
\midrule 
{\footnotesize $PC(\lambda,\mathcal{R})$}&{\footnotesize $p(\alpha)\propto\lambda e^{-\lambda \sqrt{-2\log\Gamma(\alpha)+2(\alpha-1)\psi(\alpha)-2\alpha(1-\frac{1}{\mathcal{R}})}}$}&{\scriptsize PC prior for $\alpha$ with parameter $\lambda$ and   $\mathcal{R}= \frac{\beta_1}{\beta_2}$}\\ 
{\footnotesize $Gamma(\theta)$}&{\footnotesize $p(\alpha)\propto\exp(-\theta\alpha)$}&{\scriptsize  flat prior for $\alpha$ for small $\theta $} \\ 
{\footnotesize $Gamma(\theta, \theta)$}&{\footnotesize $p(\alpha)\propto\alpha^{\theta - 1}\exp(-\theta\alpha)$}&  {\scriptsize prior for $\alpha$ with thick and thin tails $(\theta\in\{10,100\})$} \\
\midrule \midrule
{\footnotesize $SD(\epsilon)$}&{\footnotesize $p(\sigma^2)\propto\left(\frac{\sigma^2}{\epsilon}\right)^{-1/2}\exp\left(-\left(\frac{\sigma^2}{\epsilon} \right)^{1/2}\right)$}&{\tiny scale-dependent prior for $\sigma^2$}\\ 
{\footnotesize $GBP(\epsilon)$}&{\footnotesize $p(\sigma^2)\propto \left(1 + \frac{\sigma^2}{\epsilon^2} \right)^{-1}\left(\frac{\sigma^2}{\epsilon^2} \right)^{-1/2}$}&{\tiny generalised beta prime prior for $\sigma^2$ or Half-Cauchy prior for $\sigma$}\\  
{\footnotesize $G(1/2, 2\epsilon^2)$}&{\footnotesize $p(\sigma^2)\propto (\sigma^2)^{1/2-1} \exp(-\sigma^2/2\epsilon^2)$}&{\tiny gamma prior for  $\sigma^2$ or half-normal for $\sigma$}\\
{\footnotesize $IG(1,\epsilon)$}&{\footnotesize $p(\sigma^2)\propto (\sigma^2)^{-2}\exp(-\epsilon/\sigma^2)$}&{\tiny flat prior for $\sigma^2$ for $\epsilon$ is small. $\epsilon=0.005$}\\
{\footnotesize $IG(\epsilon,\epsilon)$}&{\footnotesize $p(\sigma^2)\propto (\sigma^2)^{-\epsilon - 1}\exp(-\epsilon/\sigma^2)$}&{\tiny flat prior for $\sigma^2$ on log-scale  for $\epsilon \rightarrow 0$. $\epsilon\in\{0.01, 0.001\}$ }\\
\bottomrule 
\end{tabular} 
\end{table} 
\begin{itemize} 
\item[~]\textbf{Dispersion parameter.} 
Since there is rare Bayesian research about GC distribution, we consider some priors based on its property. 
In case of flat prior, we consider $Gamma(1,\theta)$ with small $\theta$ (0.05), that was fixed in r-INLA package. The $Gamma(\theta, \theta)$ with $\theta=\{1,10,100\}$ are another alternative priors for $\alpha$ with different tails (thick and thin). Finally, we determine hyperprior for achieved PC prior in \eqref{pcalpha}. It has three hyperparameters which should be assigned. Unfortunately, principle 4 is not applicable when the rate parameter of base and flexible models are different. Therefore, we specify them due to evaluation studies. For the shrinkage parameter, $\lambda$, we consider three values as $\lambda=1, 3 and 5$. Due to \eqref{pcalpha}, we need the deviation of rate parameters of base and flexible models. Consequently, we consider some values based on its ratio as follows: $\frac{\beta_1}{\beta_2}=0.2 <1$, $\frac{\beta_1}{\beta_2}=1$ and $\frac{\beta_1}{\beta_2}=2 >1$. 
Figure \ref{figp} panel (a) shows the density of priors. 
\item[~]\textbf{Scale parameter.} 
In this case, we characterize hyperparameter of scale prior, $\tau$. 
Scale dependent prior which is obtained in \eqref{pctau}, is a type-2 Gumbel distribution with scale parameter $\theta$. To infer $\theta$ by using principle 4, $Pr(1/\sqrt{\tau}>u)=a$ that leads to $\theta=-\ln(a)/u$. Thus hyperparameters $(a,u)$ should be assigned. In this literature we consider $(a=1, u=0.01)$ (Simpson \textit{et al.}, 2017). As  it can be seen from  figure \ref{figp}, PC prior is heavy tail and concave. 
A Gamma prior with shape parameter one and scale parameter $\theta$ small as a flat prior for precision (Klein and Kneib, 2016; Gelman, 2006), is one choice and another priors are $Gamma(\theta , \theta)$ with small $\theta$. Half Cauchy as a weakly informative with the heavy tail (Gelman, 2006; Palson and Scott, 2012; Simpson \textit{et al.}, 2017), and as a final alternative prior, we consider proper uniform prior, $\tau\sim U(0, \theta)$ where $\theta$ is large. 
Figure \ref{figp} panel (b) shows the plot of density priors of $\tau$. 
\end{itemize} 
\section{Developing a Bayesian Framework}\label{Sec4} 
\subsection{Inference} 
Posterior distribution can be derived using Bayese' theorem with the likelihood from \eqref{f5} and the prior assumptions in section \ref{Sec2.3}. We complete Bayesian model formulation by the following conditional independence assumptions: 
\begin{enumerate} 
\item $\eta _{i}$, $i=1,\ldots ,n$, is the structured additive predictor of model. 
\item Priors $\pi(\bbeta_{j}|\tau_{j})$, $j=0,\ldots,J$, are conditionally independent. 
\item Priors $\pi(\alpha)$ and hyperpriors $\pi(\tau_{j})$ are mutually independent. 
\end{enumerate} 
Consequently, the joint posterior distribution of the model parameters is obtained as follows 
\begin{eqnarray} 
\label{f7} 
\pi\left(\beta_0,\bbeta, \btau,\alpha|\y\right)= 
\prod_{i=1}^{n} p(y_i|\alpha,\eta_{i})\pi(\alpha) \times \prod_{j=0}^{J}\pi(\bbeta_{j}|\tau_{j})\pi(\tau_{j}), 
\end{eqnarray} 
where $\bpsi=(\bbeta_{1}',\ldots,\bbeta_{J}')'$, $\btau=(\tau_{0},\ldots,\tau_{J})'$ and $\y=(y_1,\ldots,y_n)'$. 
The conventional approach to inference for the model \eqref{f7} is based on MCMC. It is well known, however that MCMC methods have serious problems, regarding both convergence and computational time, when applied to such models (Rue \textit{et al.}, 2009). Particularly, the complexity of the proposed model could lead to several hours or even days of computing time to implement Bayesian inference via MCMC algorithms. To overcome this issue, Rue et al. (2009) introduced the INLA method that is a deterministic algorithm and provides accurate results in seconds or minutes. INLA combines Laplace approximations (Tierney and Kadane, 1986) and numerical integration in a very efficient manner to approximate posterior marginal distributions. Let $\btheta=(\alpha, \beta_0,\tau_1,\ldots,\tau_J)^\prime$ denotes the hyperparameters of the model \eqref{f7}. Let also $\bpsi$ denotes the $pJ\times 1$ vector of latent variables. In practice, the primary interest lies in the marginal posterior distributions for each element of the latent variables vector 
\[\pi(\psi_j|\y)=\int\pi(\psi_j,\btheta|\y)d\btheta=\int\pi(\psi_j|\btheta,\y)\pi(\btheta|\y)d\btheta,~~j=1,\ldots,s,\] 
and for each element of the hyperparameter vector 
\[\pi(\theta_k|\y)=\int\pi(\btheta|\y)d\btheta_{-k},~~k=1,2,\] 
where $\btheta_{-k}$ is equal to $\btheta$ with eliminated $k$th element. The essential feature of INLA is to use this form to construct nested approximations 
\begin{eqnarray*} 
\tilde{\pi}(\psi_j|\y)&=&\int\tilde{\pi}(\psi_j|\btheta,\y)\tilde{\pi}(\btheta|\y)d\btheta, \\ 
\tilde{\pi}(\theta_k|\y)&=&\int\tilde{\pi}(\btheta|\y)d\btheta_{-k}, 
\end{eqnarray*} 
where Laplace approximation is applied to carry out the integrations required for evaluation of $\tilde{\pi}(\psi_j|\btheta,\y)$. 
A crucial success of INLA is its ability to compute model comparison criteria, such as deviance information criterion (DIC; Spiegelhalter et al., 2002) and Watanabe–Akaike information criterion (WAIC; Watanabe, 2010; Gelman et al. 2014), and various predictive measures, e.g., conditional predictive ordinate (CPO; Pettit, 1990) and probability integral transform (PIT; Dawid, 1984), to compare different possible models. Our proposed GC model has already implemented in the \texttt{R-INLA} package as a \texttt{family} argument with the name "\texttt{gammacount}". 
\subsection{Propriety of Posterior} 
One of the most important questions is whether the joint posterior distribution is proper. 
Propriety of the posterior in distributional regression 
can be ensured when combining the assumptions considered by Klein, Kneib and Lang (2015). 
They found sufficient conditions for the propriety in the general framework of structured additive distributional regression model under inverse Gamma distribution for the smoothing variance. 
Their results are based on the work of Sun \textit{et al.} 
(2001) who derived several upper and lower bounds for the required integrals. Also, Klein and Kneib (2016) 
established adapted bounds that generalized the results of 
Klein, Kneib and Lang (2015) to distributional regression with scale-dependent hyperpriors. 
They introduced some conditions for scale-dependent hyperpriors to be proper. Furthermore, Simpson \textit{et al.} (2017) showed that posterior propriety with PC prior is automatically guaranteed. Therefore, the posterior distribution in \eqref{f7} is proper forever (For more details, see Klein and Kneib, 2016; Simpson \textit{et al.}, 2017). 
%
%
%
%
\section{Experimental Assessment }\label{Sec5} 
In this section we performed several simulations in which we assessed the performance of the PC and scale dependent priors for proposed Bayesian GCSA model 
in comparison to other alternative priors discussed in Section \ref{Sec2.3}. We consider three scenarios of under-, equal- and over-dispersion for analyzing different models. 
\subsection{Simulation Settings} 
In all the scenarios, we keep simulation settings fixed to allow for a consistent comparison: 
\begin{itemize} 
\item \textit{Sample size n are chosen from $\{50,100,500\}$.} 
\item \textit{Covariates are simulated from U(-3,3) and have been centralized.} 
\item 
\textit{Simulated functional effects shown in Figure \ref{ffx}:} 
\begin{itemize} 
\item[\textbf{1}:] $f_1(x)=\sin(x)$. 
\item[\textbf{2}:] $f_2(x)=\exp(-\exp(5x))$. 
\item[\textbf{3}:] $f_3(x)=-0.5\sinh^{-1}(1.25\pi x)$. 
\end{itemize} 
\begin{figure}[ht] 
\begin{center} 
\includegraphics[scale=0.6]{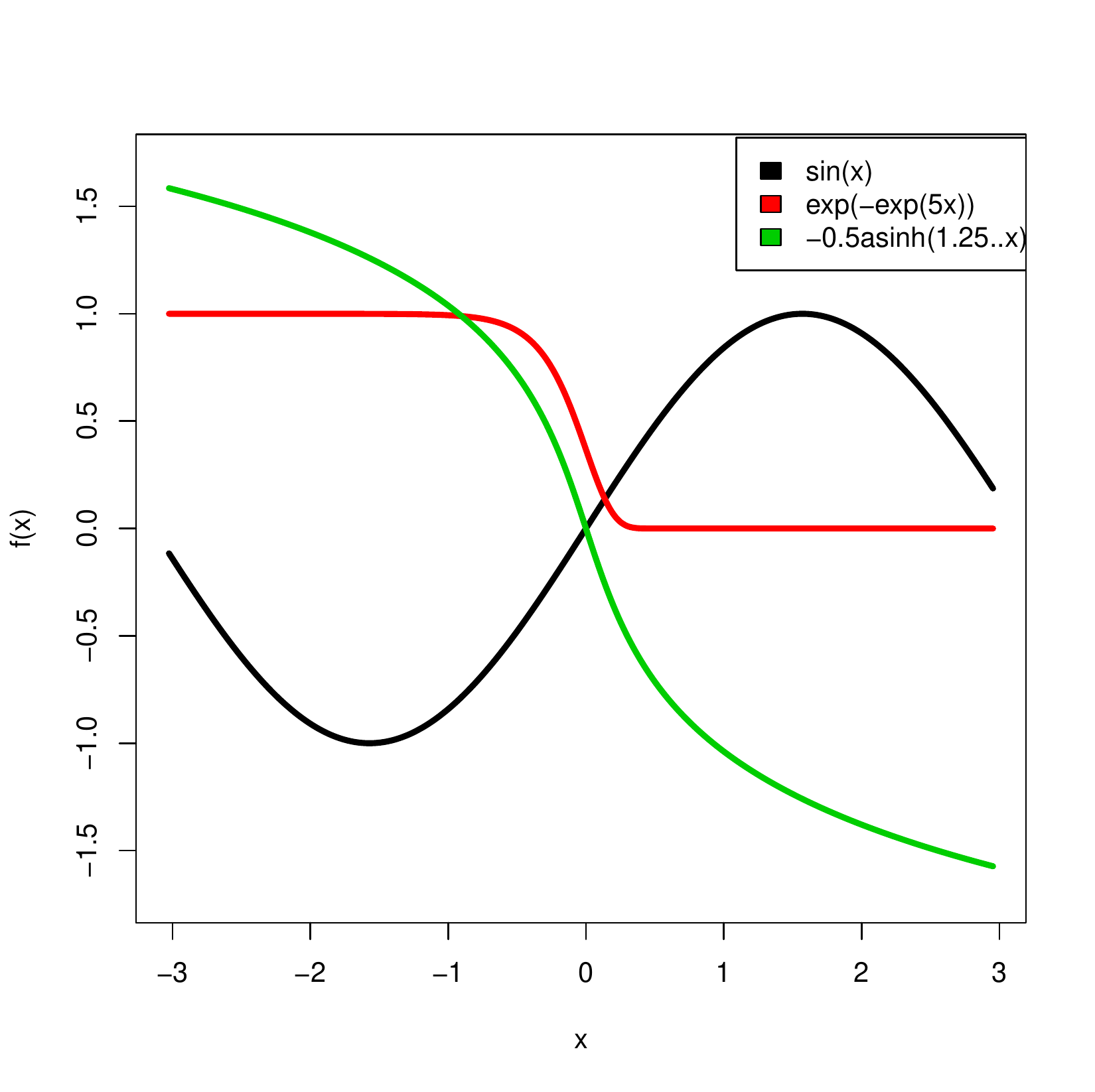} 
\caption{\label{ffx}Experimental assess: Simulated function effects $f(x)$.} 
\end{center} 
\end{figure} 
\item 
\textit{Responses are simulated as} 
\begin{center} 
$y|x \sim GC\left(\alpha, \alpha\exp\left(\beta+f(x)\right)\right)$ 
\end{center} 
\item \textit{ Hyperprior specifications} 
\begin{itemize} 
\item[$\mathbf{\alpha}$ ] 
\begin{itemize} 
\item[-] PC with equal rate parameter of base and flexible models and parameter of exponential family, $\lambda\in\{1,3,5\}$. 
\item[-] PC with rate parameter of base model equal 0.1 and flexible model equal 0.2 and parameter of exponential family, $\lambda\in\{1,3,5\}$. 
\item[-] PC with rate parameter of base model equal 0.2 and flexible model equal 0.1 and parameter of exponential family, $\lambda\in\{1,3,5\}$. 
\item[-] $(\epsilon, \epsilon)$-gamma priors with $\epsilon\in\{1,10\}$. 
\item[-] $Gamma(1, 0.005)$. 
\end{itemize} 
\item[$\mathbf{\tau}$] 
\begin{itemize} 
\item[-] scale dependent prior with PC(1,0.01). 
\item[-] $(\epsilon, \epsilon)$-gamma priors with $\epsilon\in\{0.01,0.001\}$. 
\item[-] $(1, \epsilon)$-gamma prior with $\epsilon=0.005$ as used frequently in the literature. 
\item[-] Flat prior, $Uniform(1,1000)$. 
\item[-] Half-Cauchy, $HC(0.022)$. 
\end{itemize} 
\end{itemize} 
\item \textit{Comparison criteria}\\ 
We define two criteria $Q_1$ and $Q_2$ for functional effects and $\alpha$: 
\begin{eqnarray*} 
Q_{1}(f)&=&\frac{1}{n}\sum_{i=1}^{n}(\hat{f_i}-f_i)^2\cr 
Q_{2}(\alpha)&=&(\hat{\alpha}-\alpha)^2 
\end{eqnarray*} 
where, n denotes the number of observations. 
\end{itemize} 
For each scenario, the number of replications, R, is fixed to be 500, which seems large 
enough to take into account the uncertainty in the sampling procedure. 
Also, we computed CPO as they promote the computation of the cross-validated logarithmic score. For each observation, we have 
\[{\rm CPO}_i=\pi (y_{i}|\y_{-i})~~~~~~i=1,\ldots,n,\] 
where $\y_{-i}$ is the observations vector $\y$ with the $i$th component removed. It denotes the posterior probability of observing the value of $y_i$ when the model is fitted to all data without $i$th observation. By using the values of $\{{\rm CPO}_i\}_{i=1}^{n}$, we can calculate the logarithmic score as follows: 
\[{\rm Log.score} = -\sum_{i}\log({\rm CPO}_i).\] 
A smaller value of the logarithmic score indicates a better prediction quality of the model. Finally, we computed the DIC criterion for which lower values imply better model properties. Therefore, we draw boxplot of the values of $Q_1(f)$, $Q_2(\alpha)$ and $\alpha$'s estimates. 
 
In this section, firstly we apply PC prior in equation \eqref{pcalpha} for sample size 50. According to the extraction results of this sample size, we reclaim that PC prior to different values for the rate parameters of the base and flexible model is not suitable. 
Then we apply equation \eqref{PCalpha2} for $\alpha$ under other sample sizes, 100 and 500. 
Since for there are so many scenarios and models to consider, we have included a few figures here and the rest are in the supplementary. 
\subsection{Scenario 1: Under dispersion} 
\begin{figure}[!htbp]
\begin{center} 
\begin{tabular}{c} 
\includegraphics[scale=0.3,keepaspectratio]{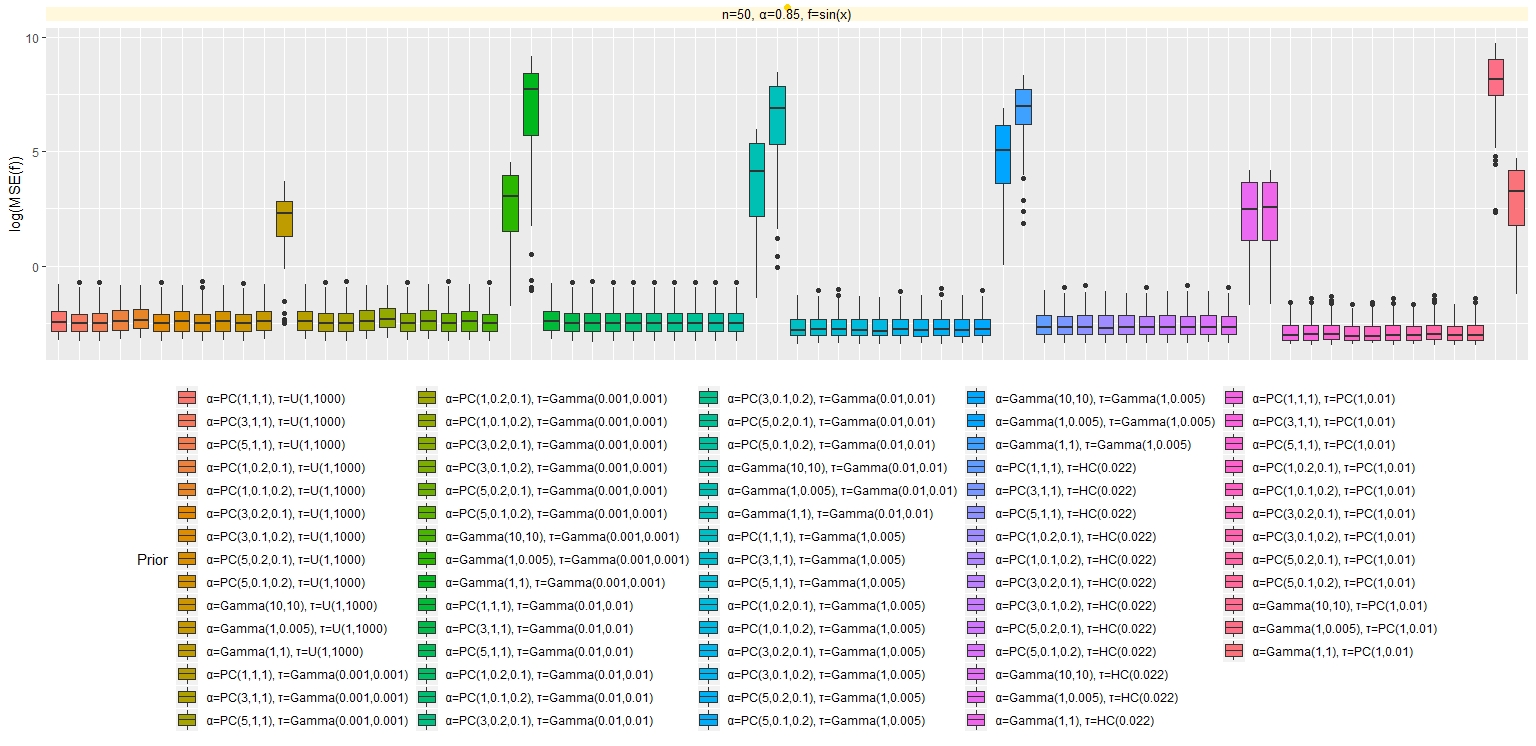} \\ 
\includegraphics[scale=0.3,keepaspectratio]{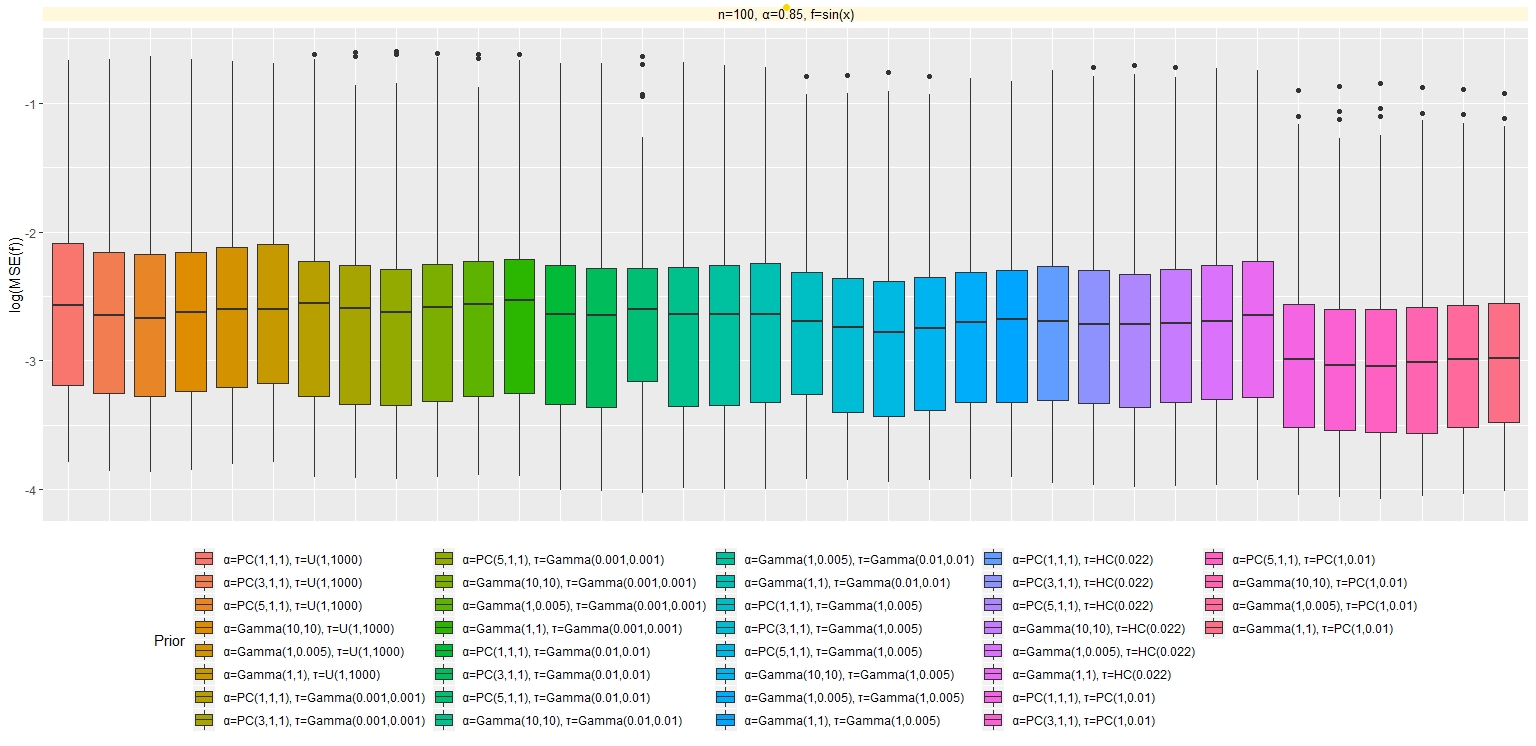} 
\end{tabular} 
\caption{\label{figp1} Experimental Assessment: The boxplot of $Q_1(f)$ values under different replications for $f=\sin(x)$ and Scenario 1.} 
\end{center} 
\end{figure}

\begin{figure}[!htbp] 
\begin{center} 
\begin{tabular}{c} 
\includegraphics[scale=0.3,keepaspectratio]{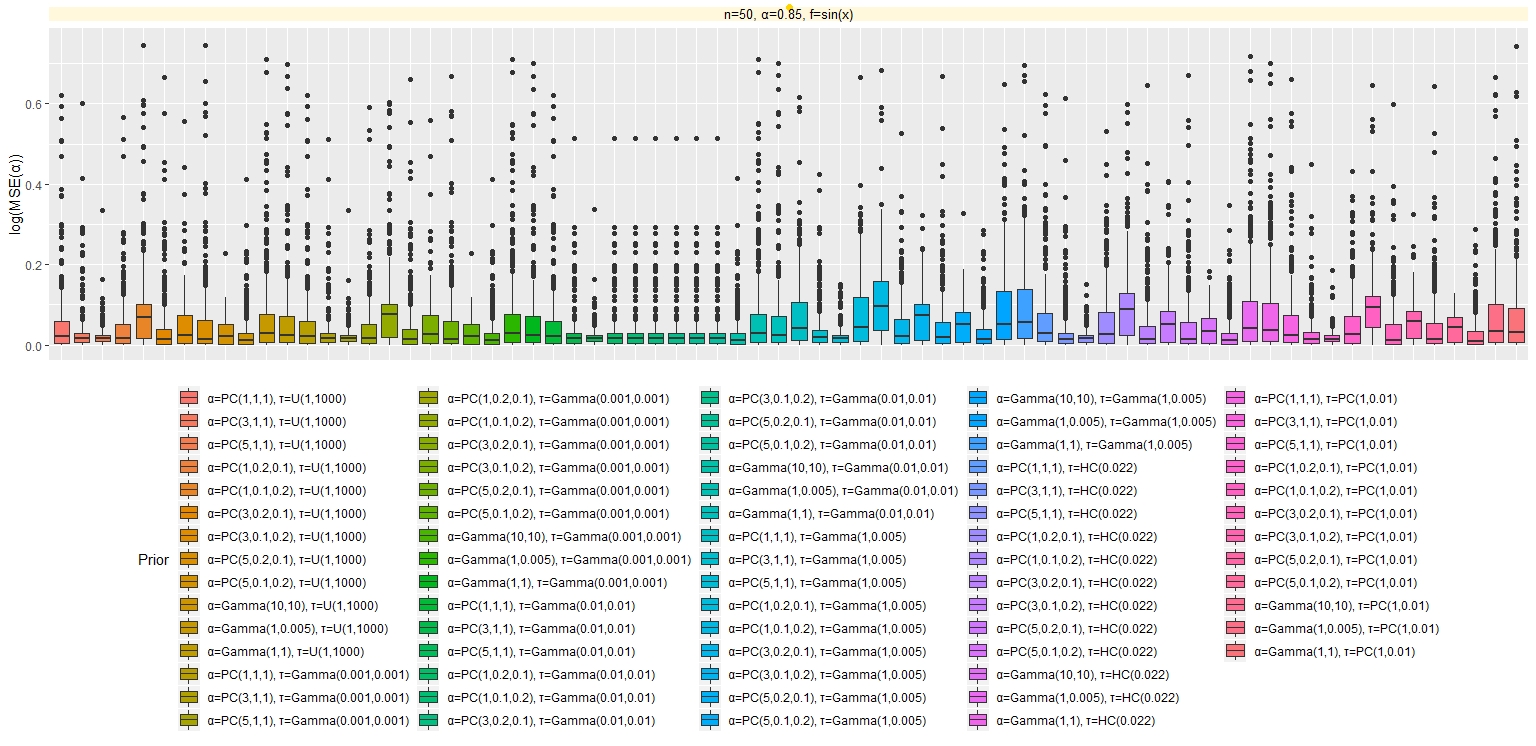} \\ 
\includegraphics[scale=0.3,keepaspectratio]{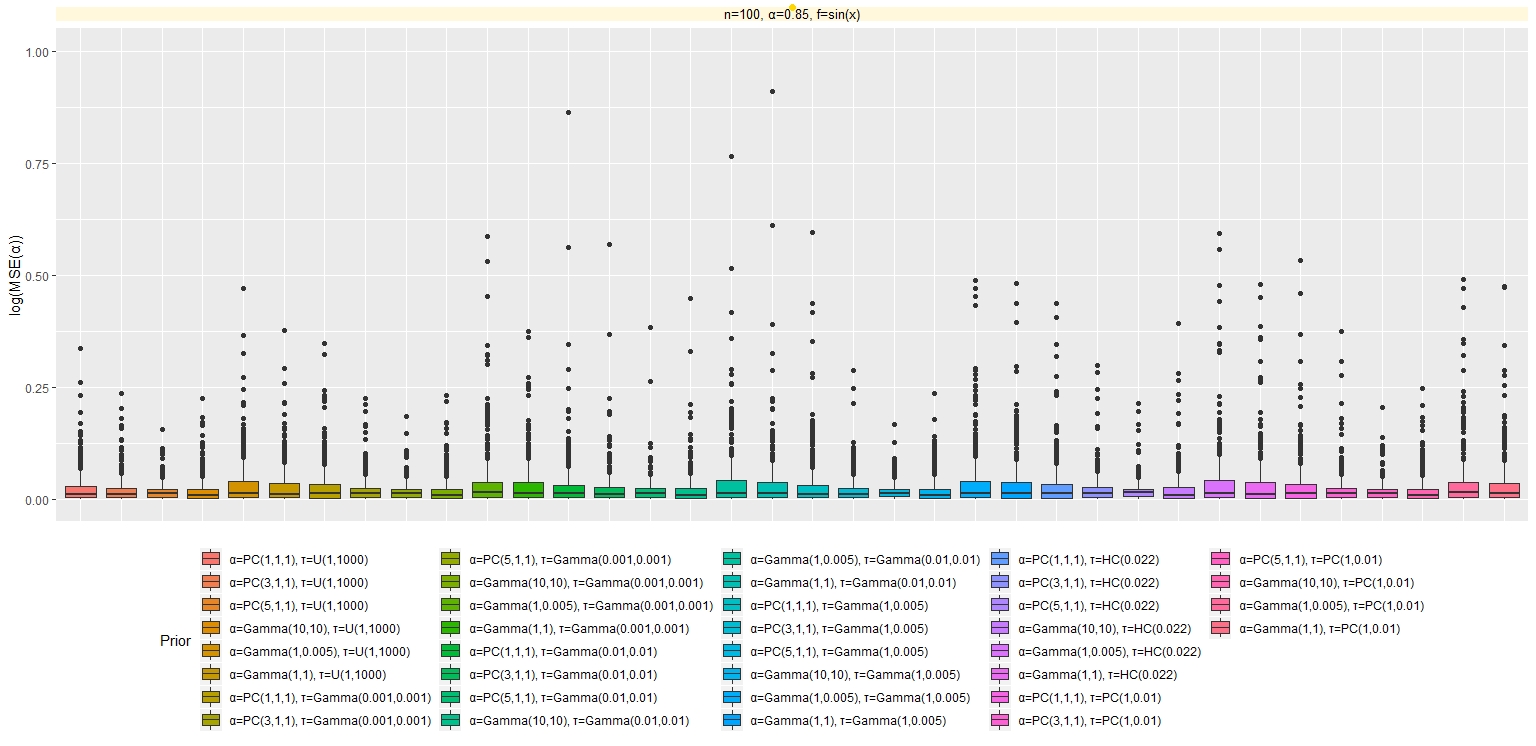} 
\end{tabular} 
\caption{\label{figp4} Experimental Assessment: The boxplot of $Q_2(\alpha)$ values under different replications for $f=\sin(x)$ and Scenario 1.} 
\end{center} 
\end{figure}

\begin{figure}[!htbp] 
\begin{center} 
\begin{tabular}{c} 
\includegraphics[scale=0.3,keepaspectratio]{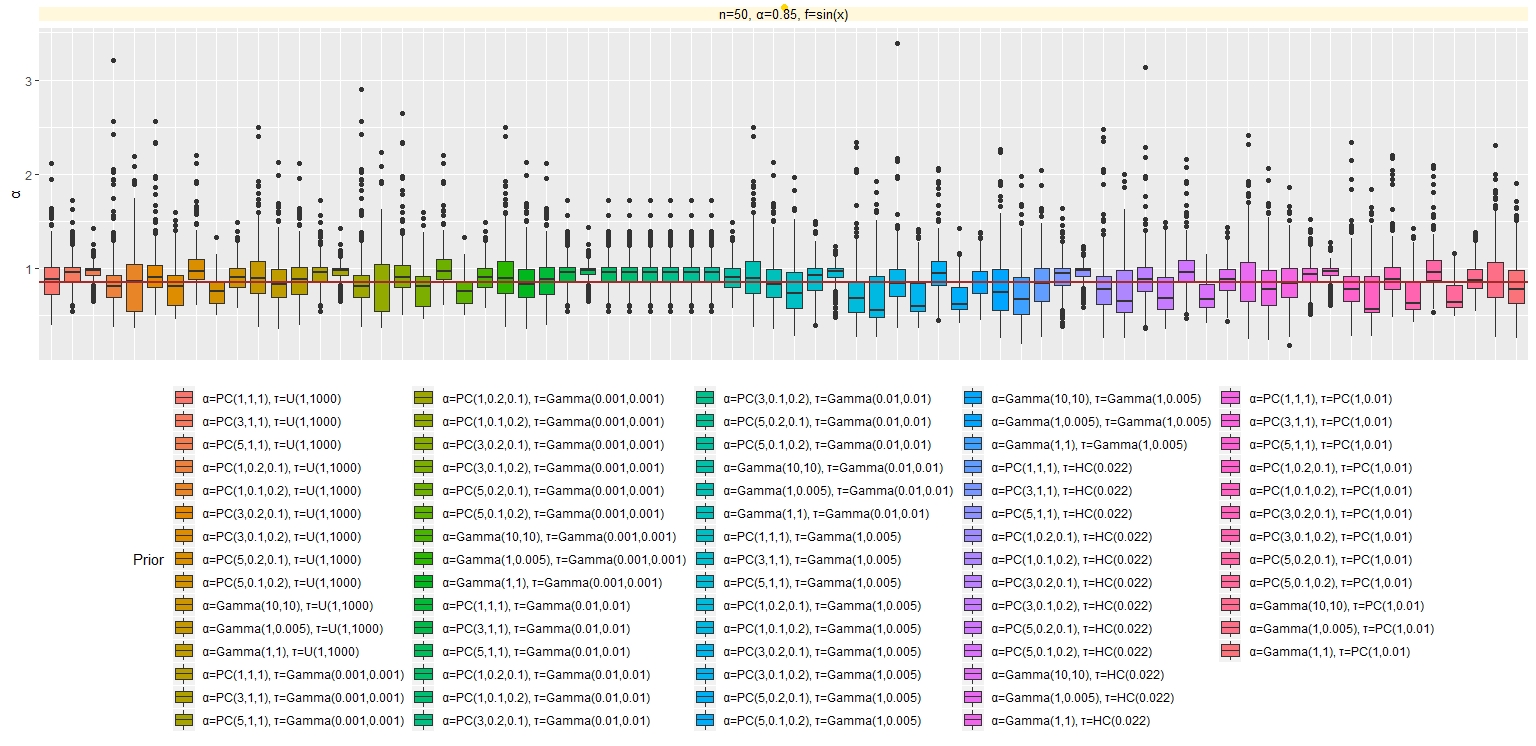} \\ 
\includegraphics[scale=0.3,keepaspectratio]{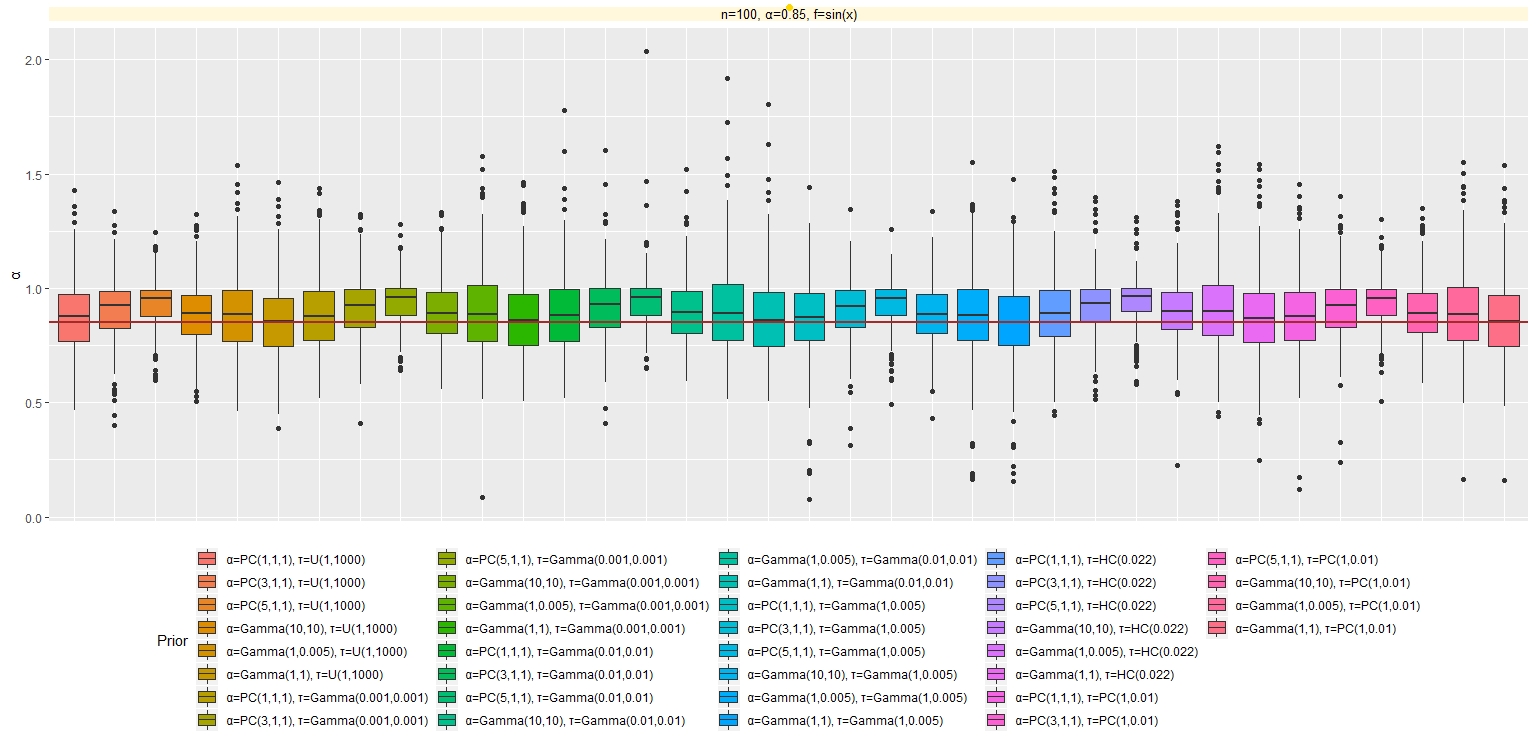} 
\end{tabular} 
\caption{\label{figp19} Experimental Assessment: The boxplot of estimated values of $\alpha$ under different replications for $f=\sin(x)$ and Scenario 1.} 
\end{center} 
\end{figure}

We observe due to the figure \ref{figp1}, except priors $Gamma(1,0.005)$ and $Gamma(1,1)$ for parameter $\alpha$, other priors have the same performance under sample size 50. Although, by increasing sample size, the likelihood dominates priors and therefore it is clear that the models have no special diversity as it can be seen from the second row of figure \ref{figp1}. Figure \ref{figp4} shows that the performance of the model corresponding to the $\tau$’s priors has the same cycle. Although, $Gamma(0.01,0.01)$ has the best efficiency among other $\tau$’s prior when the sample size is 50. As the figure \ref{figp4} exhibits amongst $\alpha$’s priors, $PC(5)$ has the best performance. After that, $PC(3)$ and $PC(1)$ are better than others, respectively. For sample size 100, this result is established, though the differences are slight. Because of $\alpha$’s estimation values in figure \ref{figp19}, the median of the estimation values of $\alpha$ under $PC(1)$, is more near the real value of $\alpha$ than $PC(5)$ or $PC(3)$. 
%
\subsection{Scenario 2: Equal dispersion} 
\begin{figure}[!htbp] 
\begin{center} 
\begin{tabular}{c} 
\includegraphics[scale=0.3,keepaspectratio]{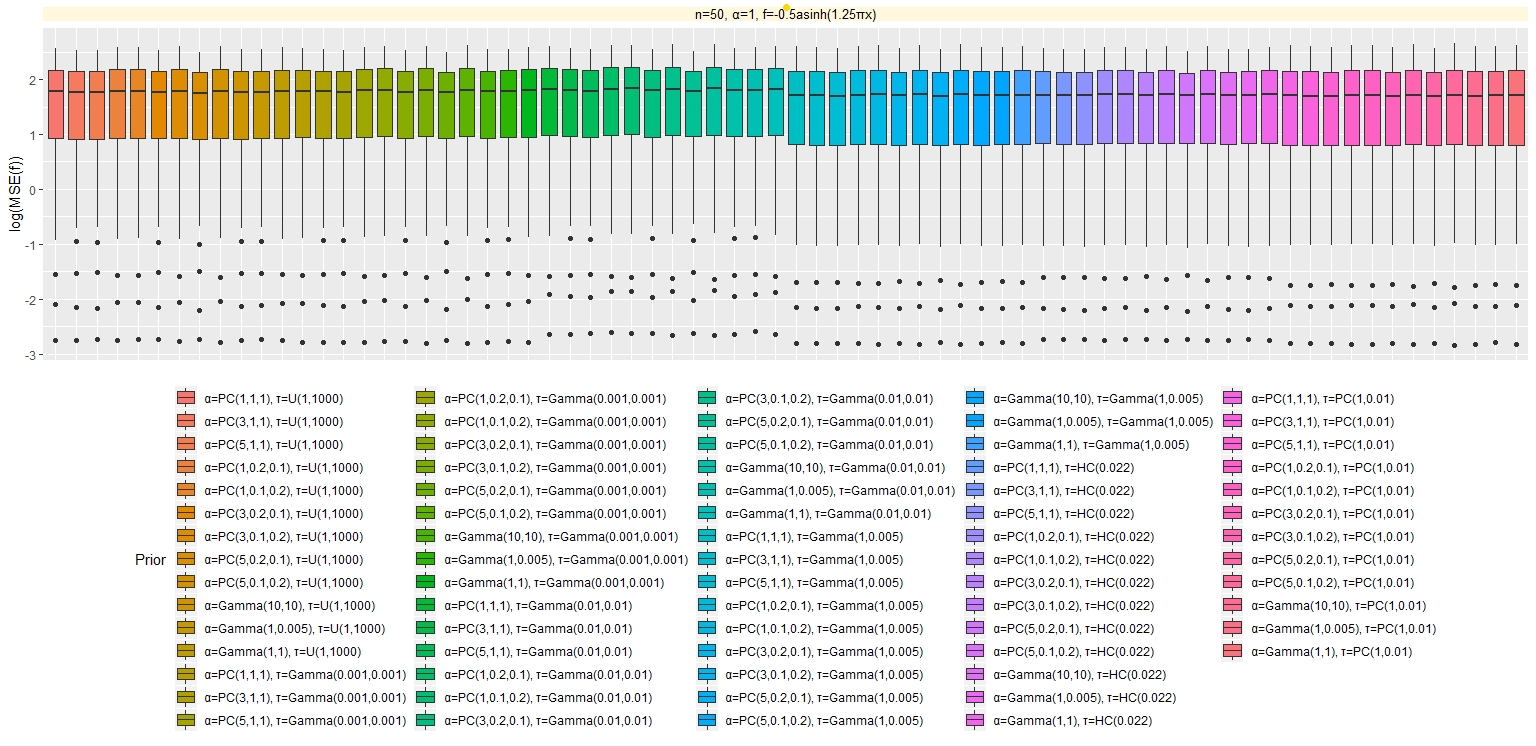} \\ 
\includegraphics[scale=0.3,keepaspectratio]{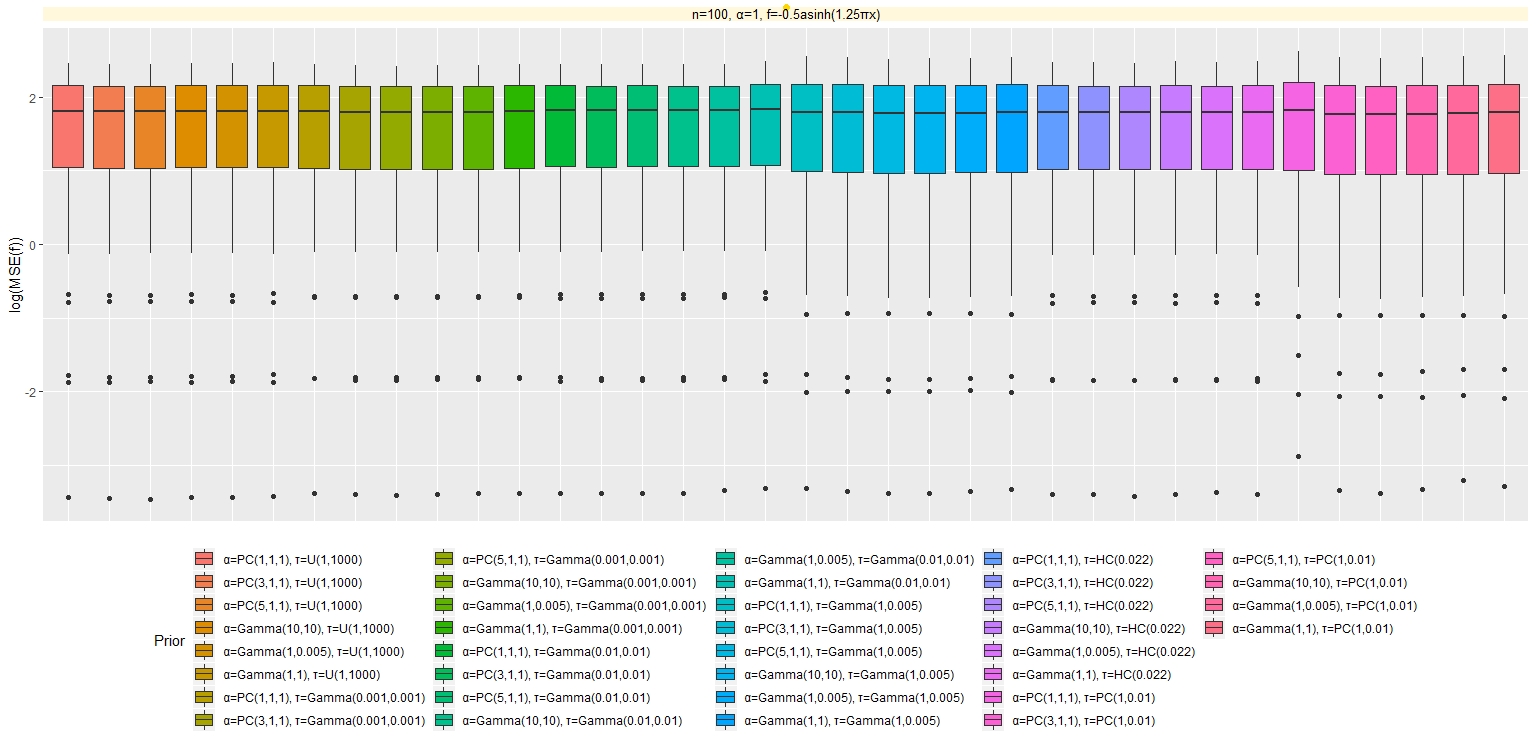} 
\end{tabular} 
\caption{\label{figp9} Experimental Assessment: The boxplot of $Q_1(f)$ values under different replications for $f=-0.5\sinh^{-1}(1.25\pi x)$ and Scenario 2.} 
\end{center} 
\end{figure}

\begin{figure}[!htbp] 
\begin{center} 
\begin{tabular}{c} 
\includegraphics[scale=0.3,keepaspectratio]{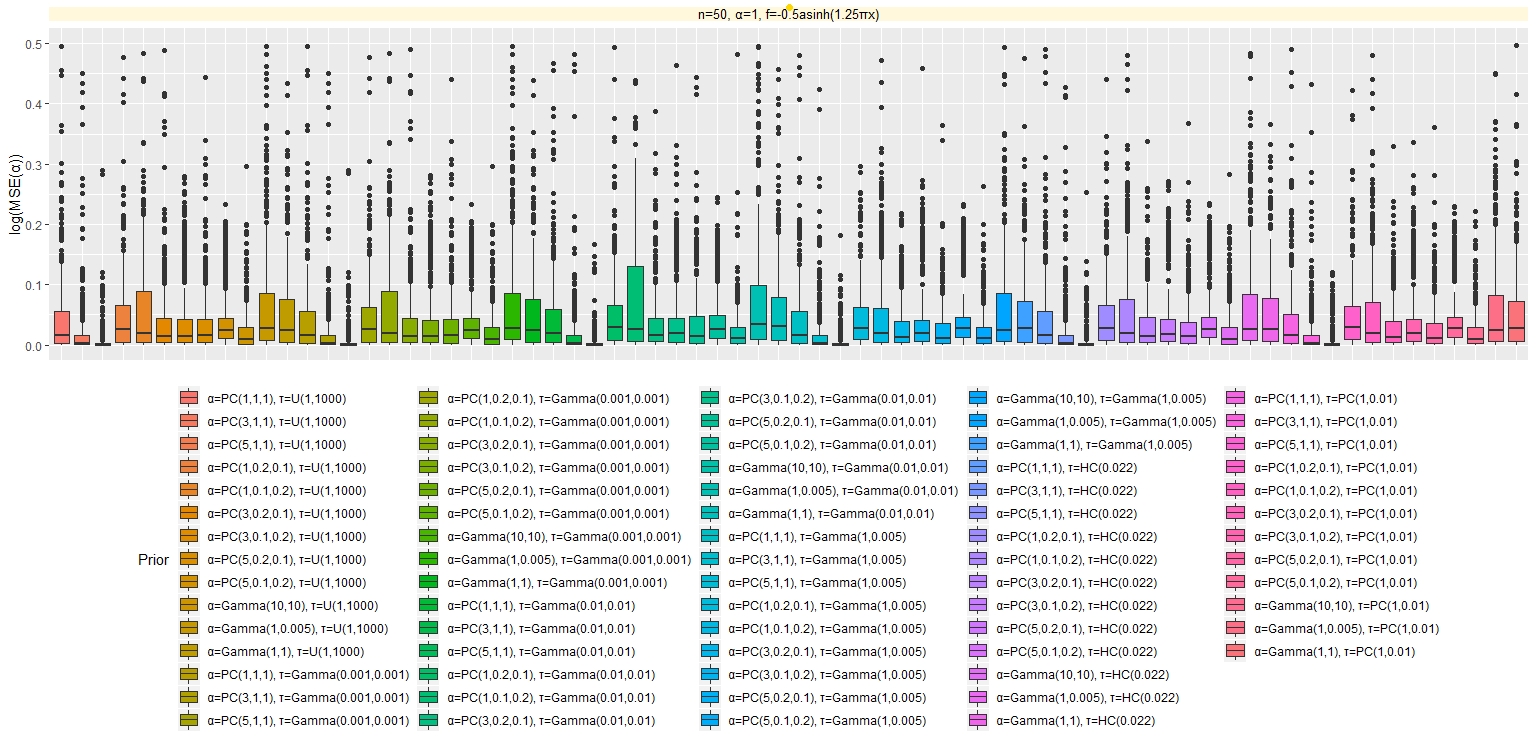} \\ 
\includegraphics[scale=0.3,keepaspectratio]{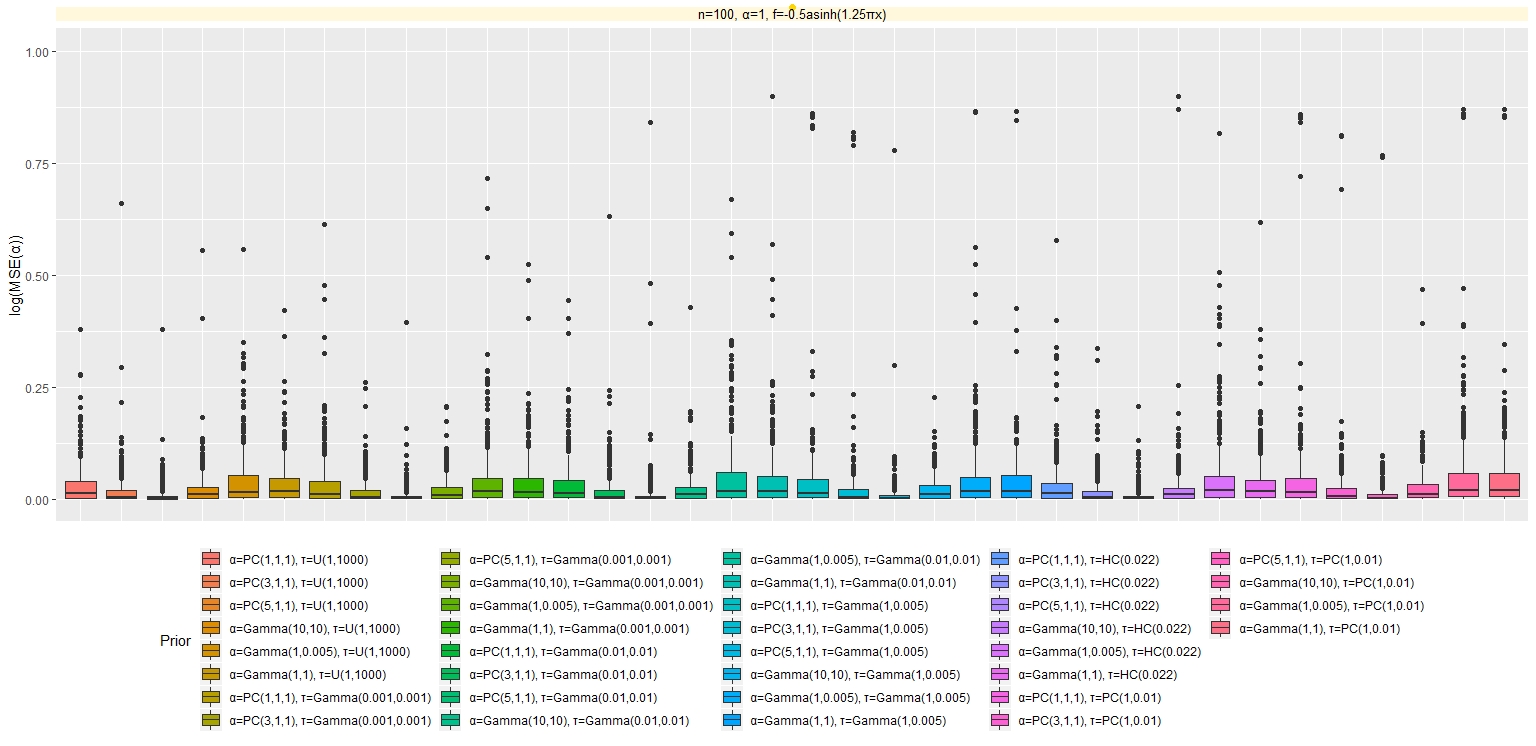} 
\end{tabular} 
\caption{\label{figp12} Experimental Assessment: The boxplot of $Q_2(\alpha)$ values under different replications for $f=-0.5\sinh^{-1}(1.25\pi x)$ and Scenario 2.} 
\end{center} 
\end{figure}

\begin{figure}[!htbp] 
\begin{center} 
\begin{tabular}{c} 
\includegraphics[scale=0.3,keepaspectratio]{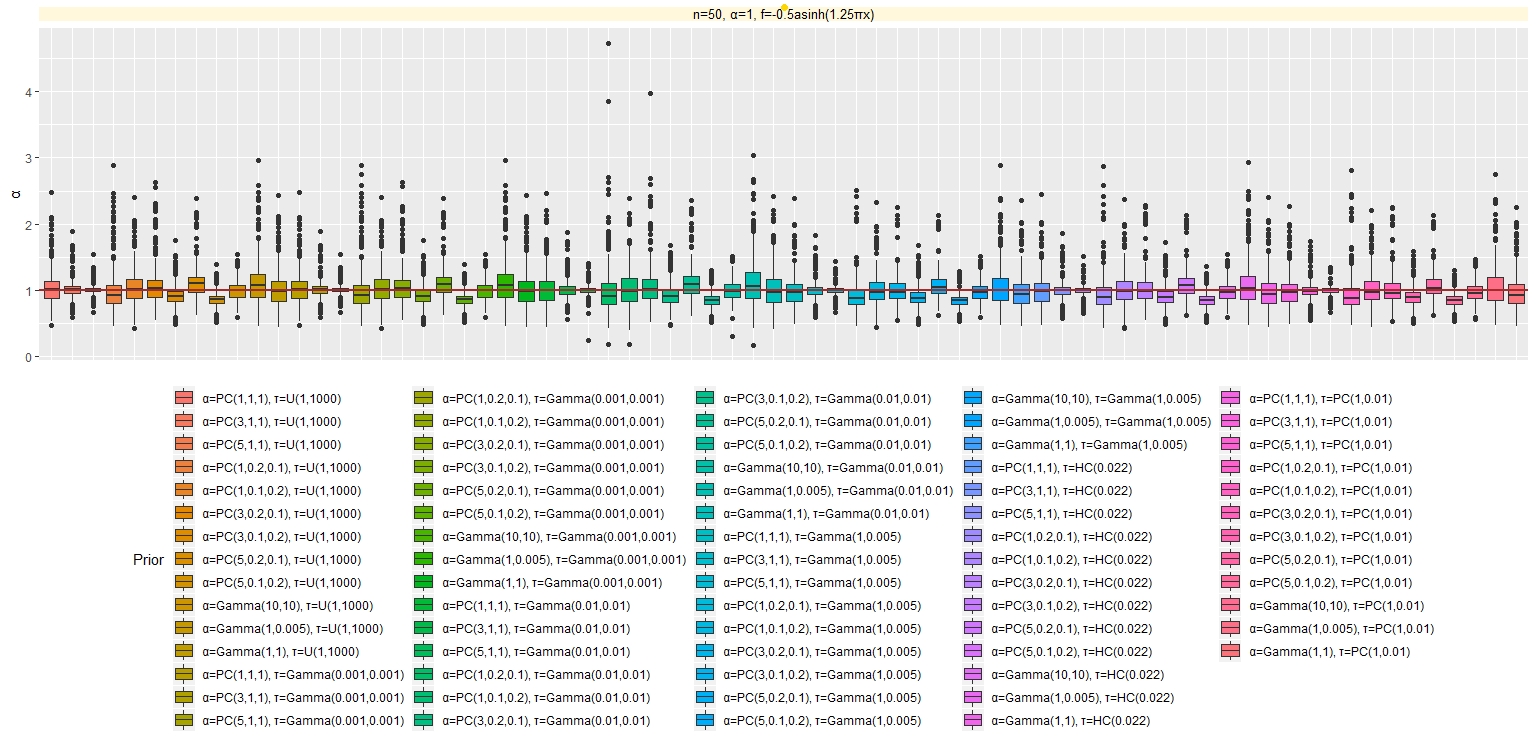} \\ 
\includegraphics[scale=0.3,keepaspectratio]{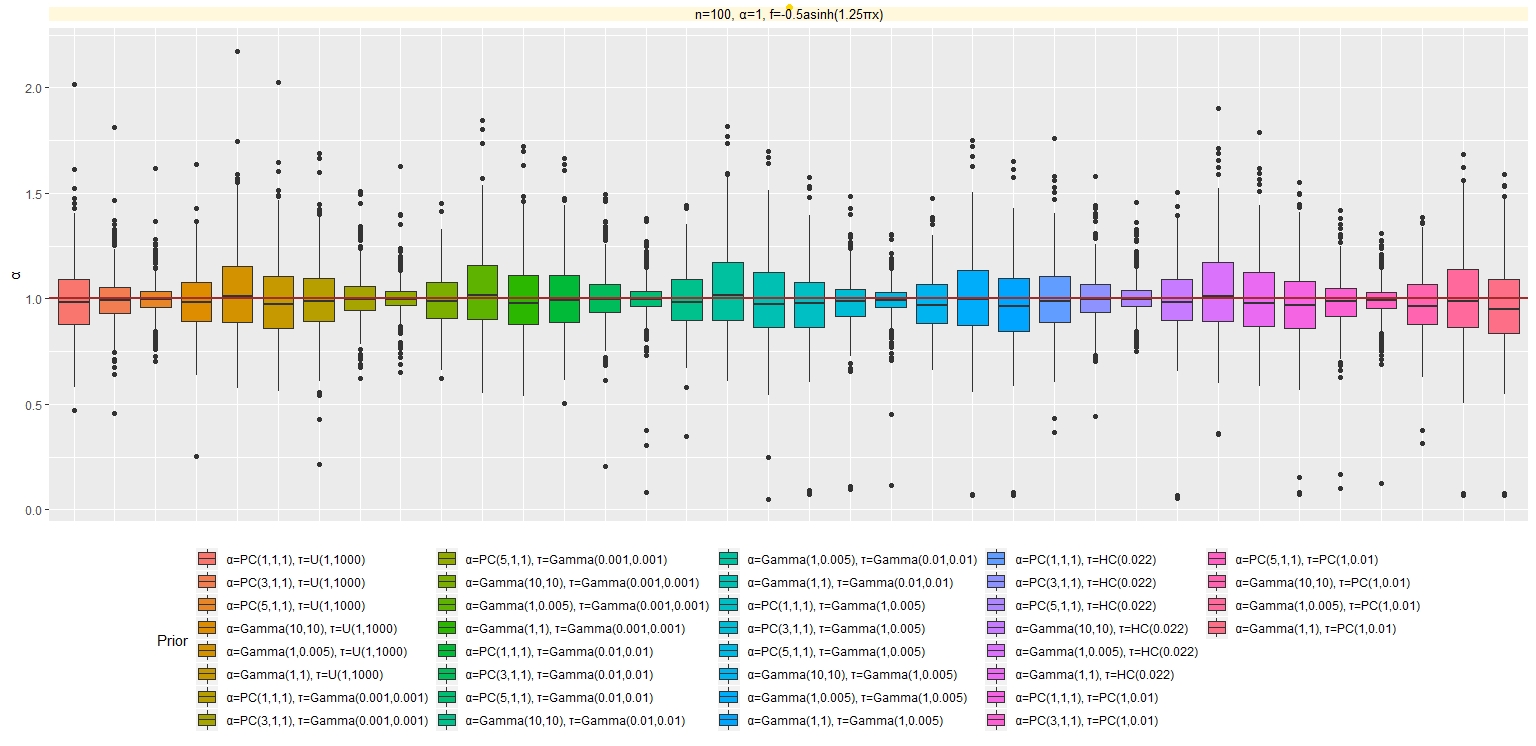} 
\end{tabular} 
\caption{\label{figp24} Experimental Assessment: The boxplot of estimated values of $\alpha$ under different replications for $f=-0.5\sinh^{-1}(1.25\pi x)$ and Scenario 2.} 
\end{center} 
\end{figure}

As it is clear from figure \ref{figp9}, all priors for parameters $\tau$ and $\alpha$ have almost the same performance , and there are little differences between the results for all sample sizes. 
Figures \ref{figp12} and \ref{figp24} present amongst $\alpha$’s priors, indicate that $PC(5)$ has the best performance. After that, $PC(3)$ and $PC(1)$ are better than others. By monitoring these figures, we can recognize when $\tau$ has the scale-dependent prior, the results are the best. 
\subsection{Scenario 3: Over dispersion} 
\begin{figure}[!htbp] 
\begin{center} 
\begin{tabular}{c} 
\includegraphics[scale=0.3,keepaspectratio]{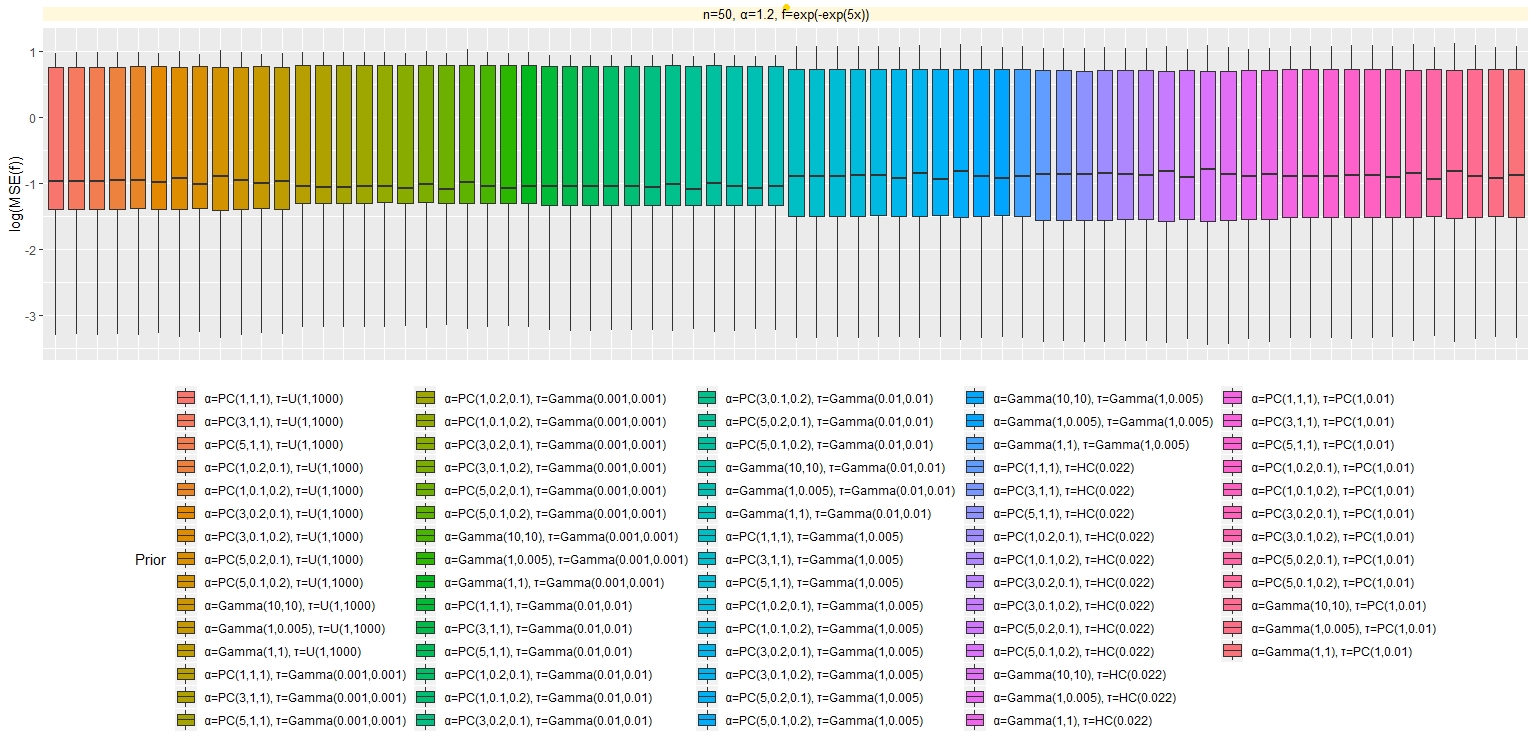} \\ 
\includegraphics[scale=0.3,keepaspectratio]{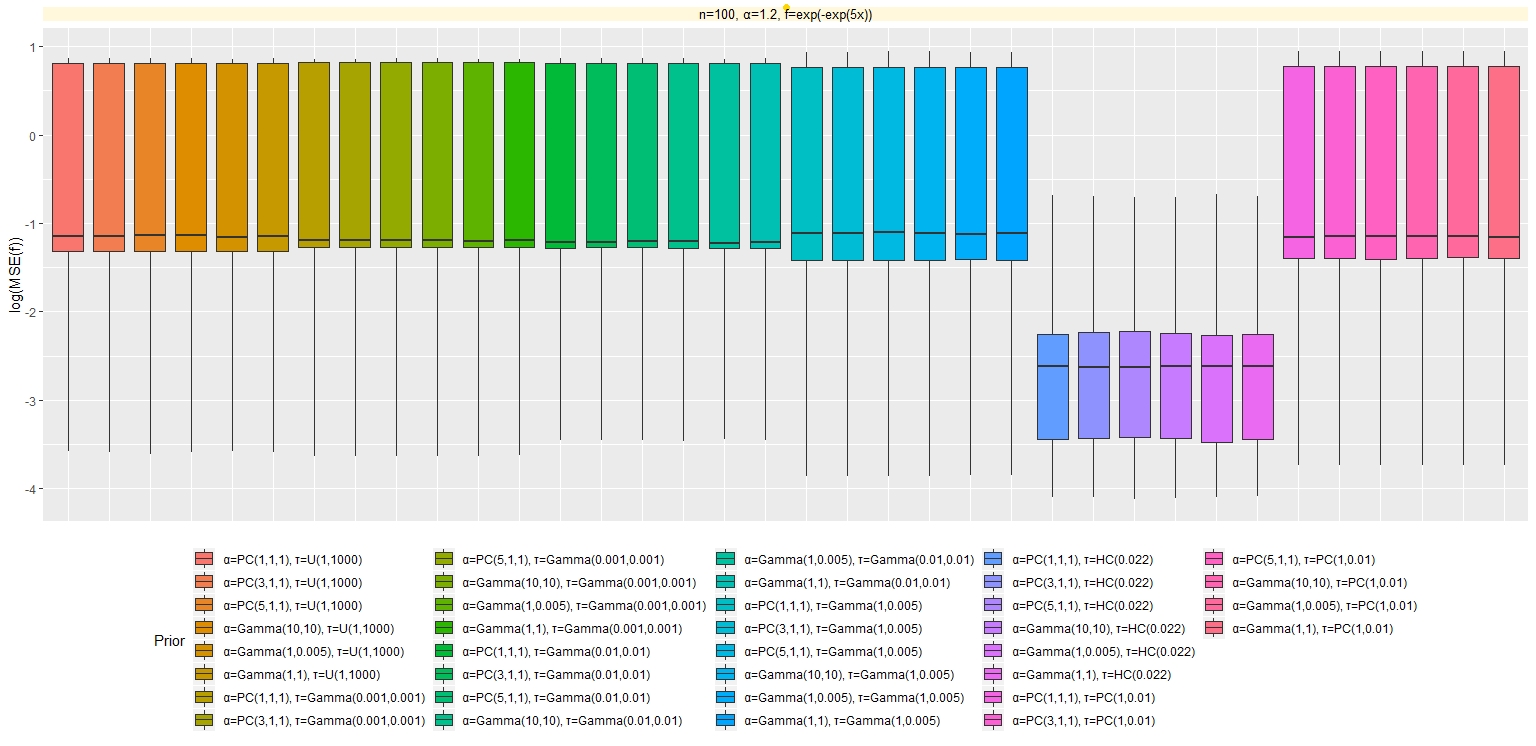} 
\end{tabular} 
\caption{\label{figp14} Experimental Assessment: The boxplot of $Q_1(f)$ values under different replications for $f=exp(-exp(5x))$ and Scenario 3.} 
\end{center} 
\end{figure}

\begin{figure}[!htbp] 
\begin{center} 
\begin{tabular}{c} 
\includegraphics[scale=0.3,keepaspectratio]{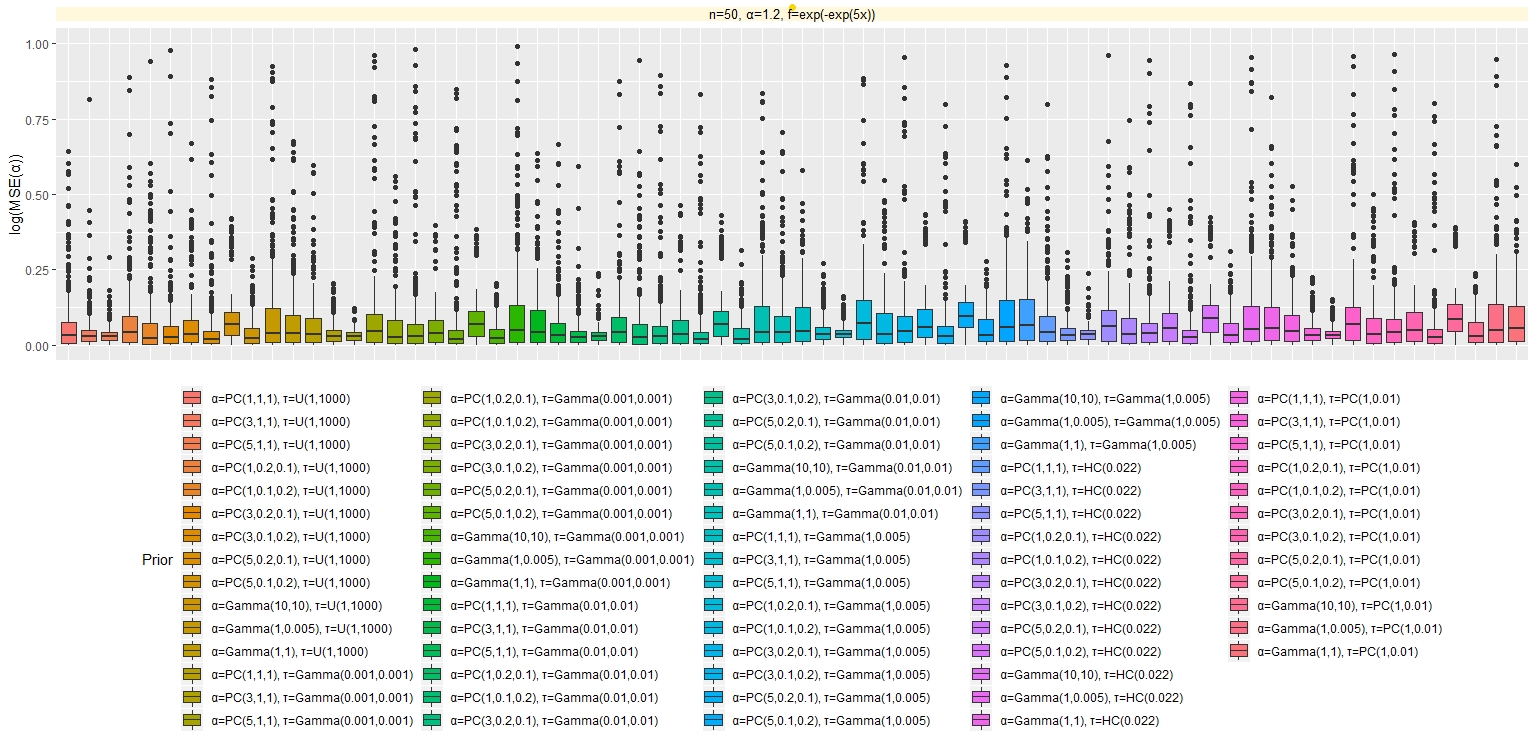} \\ 
\includegraphics[scale=0.3,keepaspectratio]{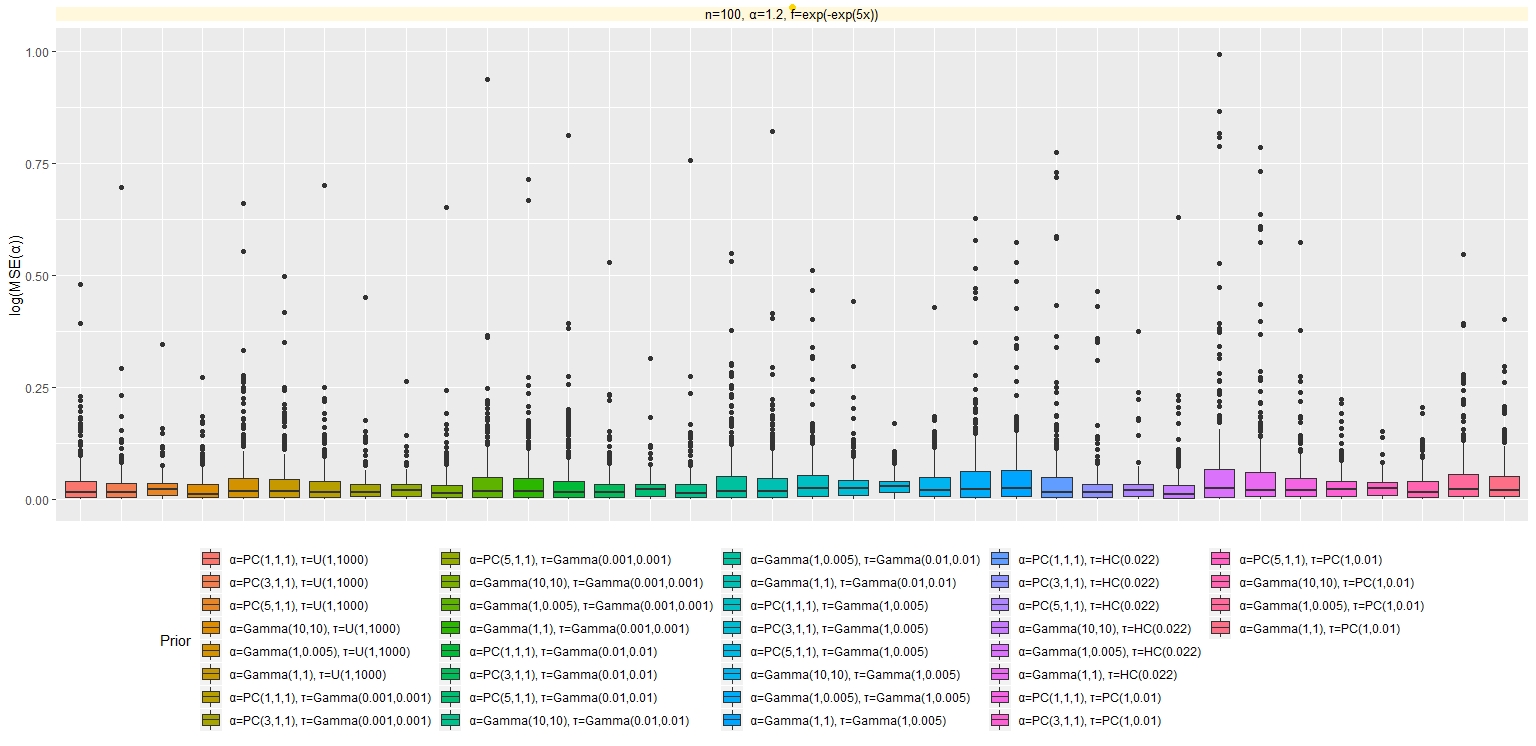} 
\end{tabular} 
\caption{\label{figp17} Experimental Assessment: The boxplot of $Q_2(\alpha)$ values under different replications for $f=exp(-exp(5x))$ and Scenario 3.} 
\end{center} 
\end{figure}

\begin{figure}[!htbp] 
\begin{center} 
\begin{tabular}{c} 
\includegraphics[scale=0.3,keepaspectratio]{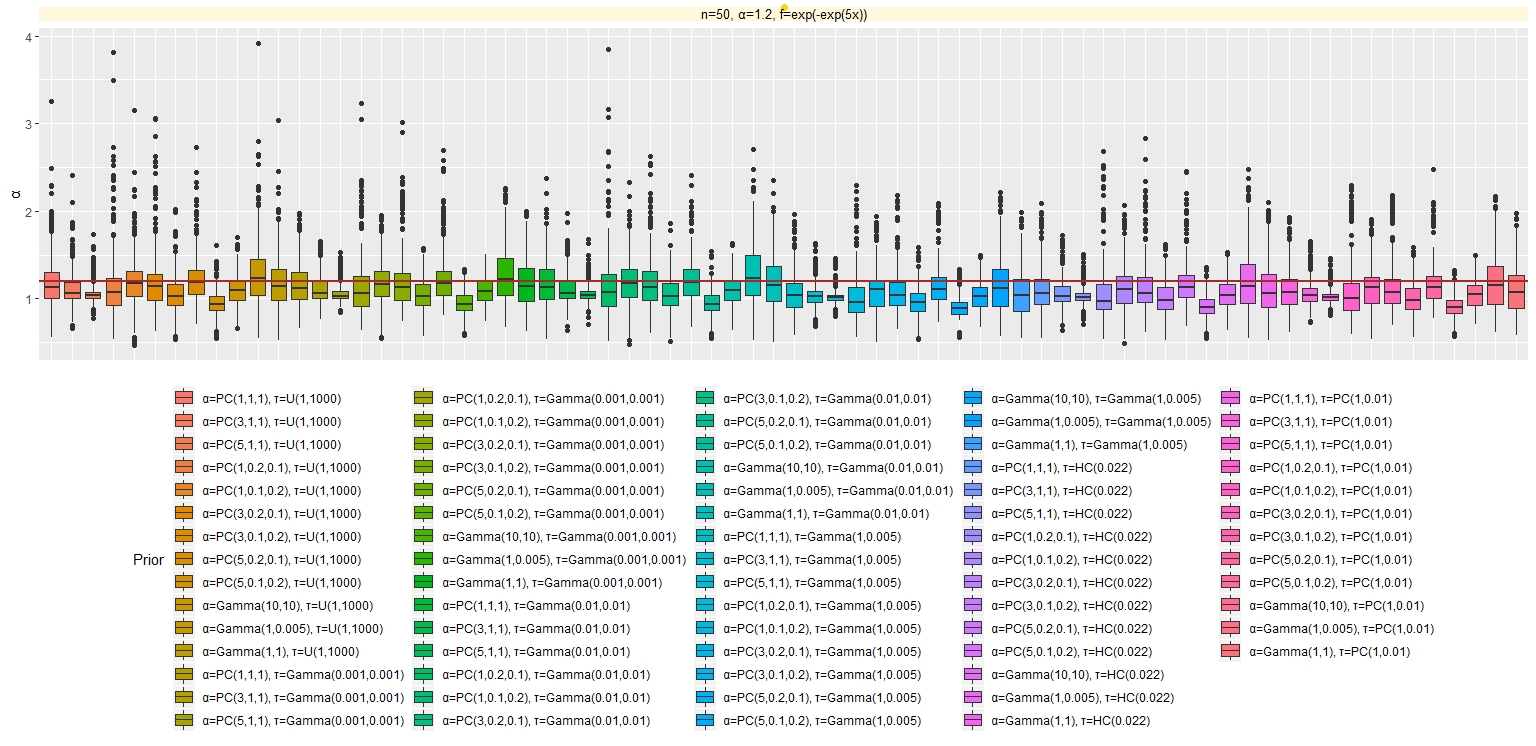} \\ 
\includegraphics[scale=0.3,keepaspectratio]{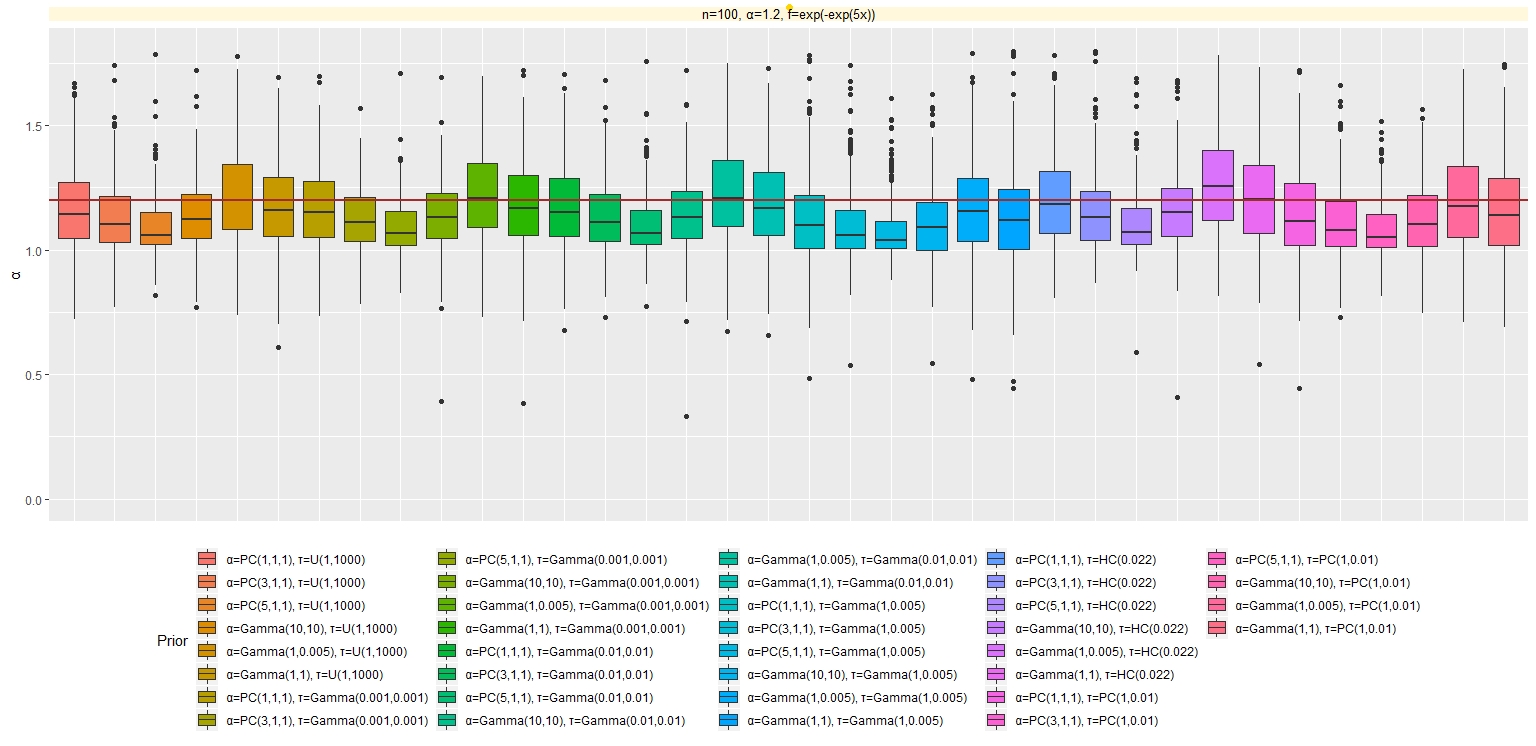} 
\end{tabular} 
\caption{\label{figp26} Experimental Assessment: The boxplot of estimated values of $\alpha$ under different replications for $f=exp(-exp(5x))$ and Scenario 3.} 
\end{center} 
\end{figure}

We can observe in figure \ref{figp14}, there are no significant differences amongst priors, and all priors has the same efficiency. 
As regards the sample size 100, there is a strange jump which $\log(MSE(f))$ values under $Half-Cauchy$ prior differ significantly from others. In the real example of handball matches you can see the same. \textcolor{red}{Why?!!!} 
Figure \ref{figp17} shows us the performance of the model corresponding to the $\tau$'s priors has the same cycle. Although, $Gamma(0.01,0.01)$ has the best efficiency among other $\tau$'s prior when sample size is 50. As the figure \ref{figp17} exhibits amongst $\alpha$'s priors, $PC(5)$ has the best performance. After that, $PC(3)$ and $PC(1)$ are better than others, respectively. There are the same results for the sample size 100, though the differences are slight. Due to the $\alpha$'s estimate values in figure \ref{figp26}, the median of the estimate values of $\alpha$ under $PC(1)$, is more near the real value of $\alpha$ than $PC(5)$ or $PC(3)$. 
%
  
\newpage
\section{Real Applications}\label{Sec6}
In this section, we briefly describe the two count data set. The first data set is related to the World Men's Handball Championships 2011 – 2017 (Groll and \textit{et al.}, 2019). As a spatial example, we consider larynx cancer mortality data in Germany, during the years of 1986 to 1990 (Nat\'{a}rio and Knorr-Held, 2003). These data sets are both under dispersed; therefore, the Poisson and the Negative Binomial (NB) models do not seem adequate for analyzing them. At first, we compare the different competitor models based on the WAIC and log score (LS) criteria. Finally, we fit the best performing model to 
the data and utilized to prediction. 
Prior distributions and their corresponding symbols are presented in the Table \ref{Tab1}. 
\begin{table}[H] 
\centering \caption{\label{Tab1} Prior distributions and their corresponding symbols.} 
\begin{tabular}{cccc} 
\multicolumn{4}{c}{Parameter}\\ 
\toprule 
\multicolumn{2}{c}{$\mathbf{\alpha}$}&\multicolumn{2}{c}{$\mathbf{\tau}$}\\ 
\cmidrule(lr){1-2}\cmidrule(lr){3-4} 
\textbf{Prior}&\textbf{Symbol}&\textbf{Prior}&\textbf{Symbol}\\ 
\midrule 
PC prior with considering $\lambda=1$&PC(l1)&PC(1,0.01)&SD\\ 
PC prior with considering $\lambda=3$&PC(l3)&Half Cauchy (0.022)& HC\\ 
PC prior with considering $\lambda=5$&PC(l5)&unif(1,1000)&Flat\\ 
Gamma(1,1)&G(1,1)&Gamma(1,0.005)&G(1,0.005)\\ 
Gamma(1,0.005)&G(1,0.005)&Gamma(0.01,0.01)&G(0.01,0.01)\\ 
Gamma(10,10)&G(10,10)&Gamma(0.001,0.001)&G(0.001,0.001)\\ 
\bottomrule 
\end{tabular} 
\end{table} 
\subsection{Handball Data} 
In this section, we consider a dataset covering all matches 
of the four preceding IHF World Men’s Handball Championships 2011 – 2017 
together with several potential influence variables. This data set are first used by Groll \textit{et al.} (2019). 
We choose the three covariates which have more measure in Groll \textit{et al.} (2019). They are observed either for the year of the respective 
World Cup or shortly before the start of it for each participating team, therefore, vary from one World Cup to 
another. 
The variables that we use in this paper are as follows. The response variable is the number of goals for each of team during the World cups. 
Both of the covariates are related to sportive factors as ODDSET probability and IHF\footnote{International Handball Federation (IHF)} ranking. 
ODDSET probability reflects the probability for each team to win the respective World Cup. Also, the IHF ranking is a ranking table of national handball federations published by the IHF (source: \url{http://ihf.info/en-us/thegame/rankingtable}). 
Another variable relates to factors of the team structure, average height. The height average of a team logically affects on the team's power because tall players might have superiority over short players, as they can release a shot on goal above a defender more. 
As already stated in previous literature, each score from a match of two handball teams is treated 
as a count observation. Accordingly, for n teams, the respective model has the form 
\begin{eqnarray} 
\label{IHF1} 
y_{ijk} &\sim & GC(\alpha , \alpha\eta_{ij}),\cr 
\log(\eta_{ijk}) &=& f(odds_{ij}) + f(rank_{ij}) + f(height_{ij}), 
\end{eqnarray} 
where $y_{ijk}$ denotes the number of goal of team $i$ versus team $j$ in tournament $k$ with $i\neq j \in\{1,\ldots,n\}$. 
Tables \ref{Tab2-1} and \ref{Tab2-2} provide LS and WAIC criteria for 
model corresponding to the different priors. 
\begin{table}[H] 
\begin{scriptsize} 
\centering \caption{\label{Tab2-1}LS criterion corresponding different models and priors for handball data} 
\begin{tabular}{rrrrrrrr} 
\toprule 
\textbf{Model}& \multicolumn{6}{c}{\textbf{Priors}}&\\ 
\cmidrule{2-8} 
&$\alpha$&\multicolumn{5}{c}{$\tau$}\\ 
\cmidrule{3-8} 
&& \textbf{SD} & \textbf{HC} & \textbf{Flat} & \textbf{G(1,0.005)} & \textbf{G(0.01,0.01)} &\textbf{G(0.001,0.001)}\\ 
\midrule 
&\textbf{PC(l1)} & 496.94 &\textbf{496.08}&497.22&503.99&506.62& 501.80\\ 
& \textbf{PC(l3)}&496.95 &496.14&497.21&503.92&506.52&501.78 \\ 
\textbf{GC}&\textbf{PC(l5)} & 497.03 &496.31&497.28&503.94&506.49&501.84 \\ 
& \textbf{G(10,10)}&496.82 &496.95&497.07&503.81&506.39& 501.62\\ 
& \textbf{G(1,0.005)}&496.97 &496.13&497.27&504.06&506.78& 501.91\\ 
& \textbf{G(1,1)}& 496.93&496.13&497.25&503.98&506.68& 501.83\\ 
\midrule 
\textbf{NB}&--& 499.27&498.85&499.68&505.70&508.24& 503.96\\ 
\textbf{Poisson}&--& 498.64&498.23&499.08&505.13&507.65& 503.40\\ 
\bottomrule 
\end{tabular} 
\end{scriptsize} 
\end{table} 
\begin{table}[H] 
\begin{scriptsize} 
\centering \caption{\label{Tab2-2} WAIC criterion corresponding different models and priors for handball data} 
\begin{tabular}{rrrrrrrr} 
\toprule 
\textbf{Model}& \multicolumn{6}{c}{\textbf{Priors}}&\\ 
\cmidrule{2-8} 
&$\alpha$&\multicolumn{5}{c}{$\tau$}\\ 
\cmidrule{3-8} 
&& \textbf{SD} & \textbf{HC} & \textbf{Flat} & \textbf{G(1,0.005)} & \textbf{G(0.01,0.01)} &\textbf{G(0.001,0.001)}\\ 
\midrule 
&\textbf{PC(l1)} & 993.71&\textbf{991.92}&993.97&1006.12&1009.57& 1001.90\\ 
& \textbf{PC(l3)}&993.75 &992.07&994.01&1006.09&1009.57& 1001.94 \\ 
\textbf{GC}&\textbf{PC(l5)} & 993.9172 &992.41&994.19&1006.24&1009.73&1002.16 \\ 
& \textbf{G(10,10)}& 993.49&991.96&993.72&1005.84&1009.28& 1001.61\\ 
& \textbf{G(1,0.005)}&993.76 &992.01&994.05&1006.20&1009.69& 1002.03\\ 
& \textbf{G(1,1)}& 993.68&991.97&994.02&1006.08&1009.60& 1001.92\\ 
\midrule 
\textbf{NB}&-- &998.45&997.60&999.14&1010.26&1014.27& 1006.91\\ 
\textbf{Poisson}& --&997.18&996.36&997.93&1009.05&1012.99& 1005.74\\ 
\bottomrule 
\end{tabular} 
\end{scriptsize} 
\end{table} 
We can conclude some points according Tabels \ref{Tab2-1} and \ref{Tab2-2}. 
GC model with priors $PC(l1)$ for $\alpha$ and $HC$ for $\tau$ represents a better fit than others in terms of WAIC and LS (logarithmic score) criteria. 
\subsection{Bym: Larynx Cancer} 
\begin{figure}[!htbp] 
\begin{center} 
\begin{tabular}{ccc} 
\includegraphics[scale=0.2]{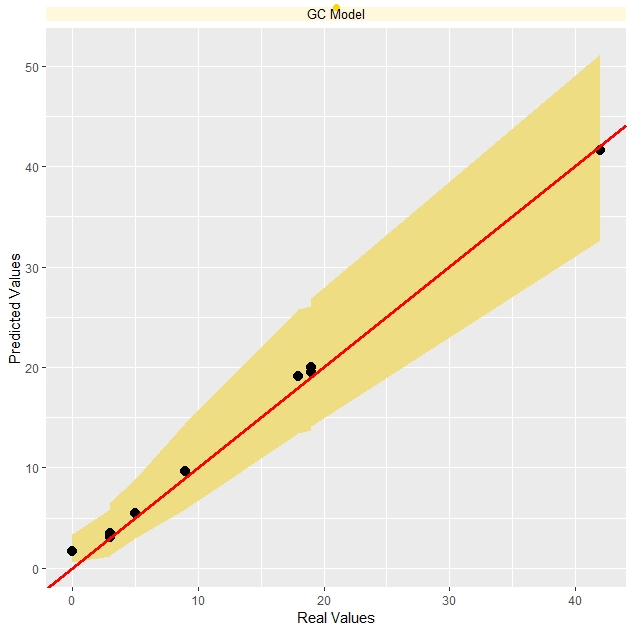} & \includegraphics[scale=0.2]{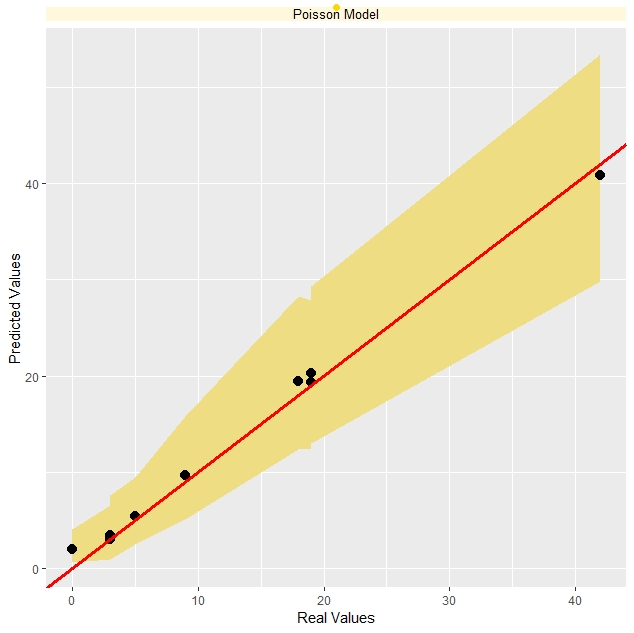}&\includegraphics[scale=0.2]{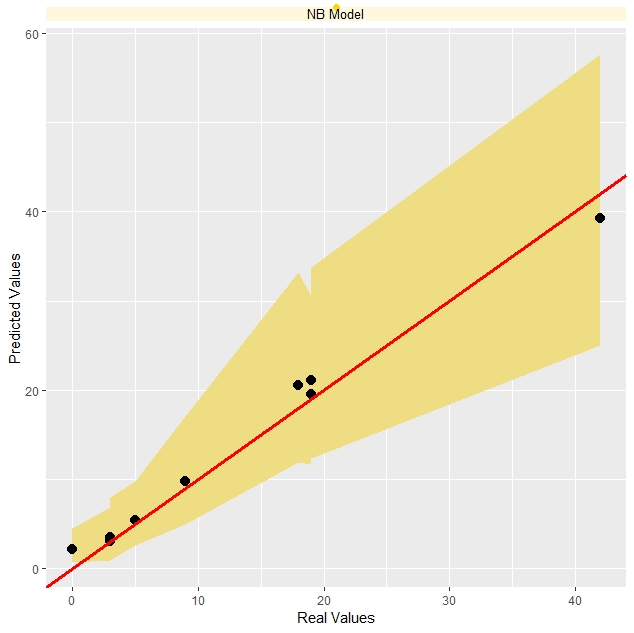}\\ 
\end{tabular} 
\caption{\label{fig1} \footnotesize{Scatter plots of observed versus predicted values for the hold-out districts and $95\%$ credible intervals with 1:1 line under the GC (first column), Poisson (second column) and NB (third column) models.}} 
\end{center} 
\end{figure}

\begin{figure}[!htbp] 
\begin{center} 
\begin{tabular}{ccc} 
\includegraphics[scale=0.2]{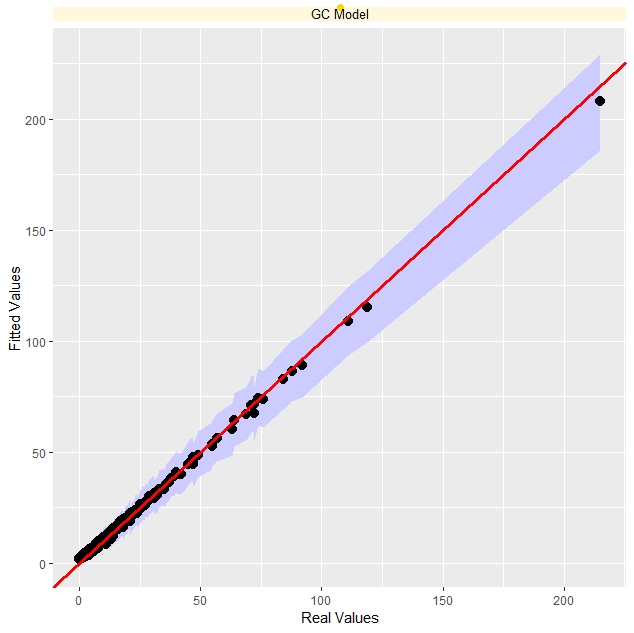} & \includegraphics[scale=0.2]{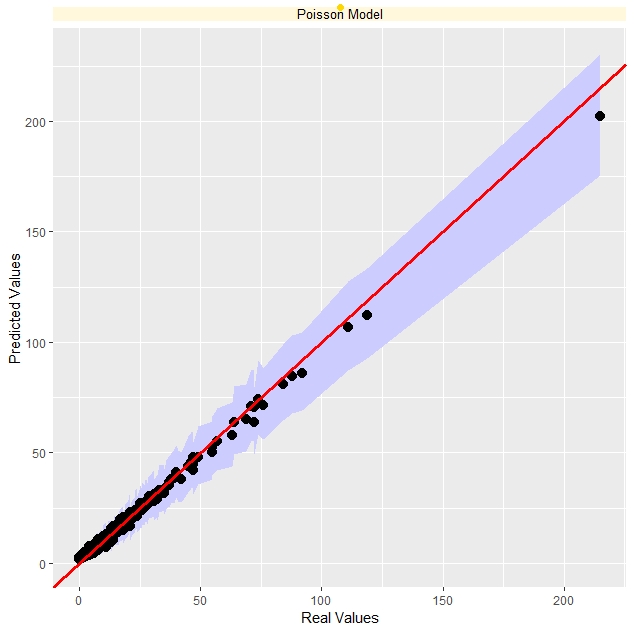}&\includegraphics[scale=0.2]{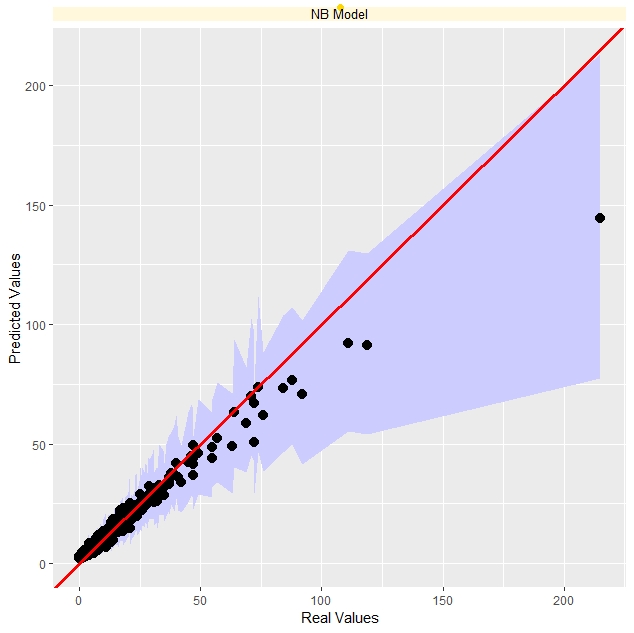}\\ 
\end{tabular} 
\caption{\label{fig2} \footnotesize{Scatter plots of observed versus fitted values for the hold-out districts and $95\%$ credible intervals with 1:1 line under the GC (first column), Poisson (second column) and NB (third column) models.}} 
\end{center} 
\end{figure}

Here, we reconsider larynx cancer mortality data of Nat\'{a}rio and Knorr-Held (2003) for males, in the 544 districts of Germany, from 1986 to 1990. During this period 7,283 deaths were recorded due to larynx cancer among the male population. Nat\'{a}rio and Knorr-Held (2003) used the information on the corresponding lung cancer mortality rates in the same period, as an ecological covariate to account for smoking consumption $c$. This dataset is available on the INLA website and see Nat\'{a}rio and Knorr-Held (2003) for more details. 
We assume the observed death counts $y_i$ in district $i=1,\ldots,544,$ are conditionally independent with a count distribution for which $\eta_i$ is given by 
\[\eta_i=\beta_0 + \beta_1 c_i+f_i(s),\] 
where $f_i(s)$ is an ICAR spatial effect, as described in Section \ref{secMRF}. We fit the GC, Poisson, and NB regression models as the count distribution of the responses under priors described in previous sections. 
Tables \ref{Tab3} and \ref{Tab4} provide LS and WAIC criteria for model corresponding to the described priors. 
\begin{table}[H] 
\begin{scriptsize} 
\centering \caption{\label{Tab3} LS criterion corresponding different models and priors for larynx cancer} 
\begin{tabular}{rrrrrrrr} 
\toprule 
\textbf{Model}& \multicolumn{6}{c}{\textbf{Priors}}&\\ 
\cmidrule{2-8} 
&$\alpha$&\multicolumn{5}{c}{$\tau$}\\ 
\cmidrule{3-8} 
&& \textbf{SD} & \textbf{HC} & \textbf{Flat} & \textbf{G(1,0.005)} & \textbf{G(0.01,0.01)} &\textbf{G(0.001,0.001)}\\ 
\midrule 
&\textbf{PC(l1)} & \textbf{1936.46} & 2313.67 & 1952.95 & 2043.67 & \textbf{1926.24} & 1999.19 \\ 
& \textbf{PC(l3)}& 1961.98 & 1963.85 & 1964.48 & 1964.11 & 1964.47 & 1964.59 \\ 
\textbf{GC}&\textbf{PC(l5)} & 1977.24 & 1981.80 & 1981.52 & 1980.23 & 1980.66 & 1981.61 \\ 
& \textbf{G(10,10)} & 1966.08 & 1969.26 & 1969.77 & 1969.16 & 1969.59 & 1969.54 \\ 
& \textbf{G(1,0.005)} & 1977.39 & 1976.86 & 2012.56 & 2221.87 & 2120.75 & 4854.53 \\ 
& \textbf{G(1,1)}& 2246.78 & 2103.87 & 1972.53 & 2079.13 & 2438.30 & 4905.43 \\ 
\midrule 
\textbf{NB}&-- & 11744.78 & 16920.37 & 16146.39 & 17627.93 & 16262.42 & 16168.92 \\ 
\textbf{Poisson}& --& 1850.74 & 1855.99 & 1856.46 & 1855.41 & 1856.55 & 1856.54 \\ 
\bottomrule 
\end{tabular} 
\end{scriptsize} 
\end{table} 
\begin{table}[H] 
\begin{scriptsize} 
\centering \caption{\label{Tab4} WAIC criterion corresponding different models and priors for larynx cancer} 
\begin{tabular}{rrrrrrrr} 
\toprule 
\textbf{Model}& \multicolumn{6}{c}{\textbf{Priors}}&\\ 
\cmidrule{2-8} 
&$\alpha$&\multicolumn{5}{c}{$\tau$}\\ 
\cmidrule{3-8} 
&& \textbf{SD} & \textbf{HC} & \textbf{Flat} & \textbf{G(1,0.005)} & \textbf{G(0.01,0.01)} &\textbf{G(0.001,0.001)}\\ 
\midrule 
&\textbf{PC(l1)} & \textbf{2793.59} & 17523.51 & 2838.71 & 3027.33 & \textbf{2744.78} & 2948.04 \\ 
& \textbf{PC(l3)} & 2854.951 & 2839.25 & 2837.35 & 2840.74 & 2837.53 & 2837.55 \\ 
\textbf{GC} &\textbf{PC(l5)} & 2905.59 & 2899.97 & 2898.99 & 2899.44 & 2897.50 & 2898.96 \\ 
& \textbf{G(10,10)} & 2876.99 & 2867.71 & 2866.38 & 2868.50 & 2866.54 & 2866.65 \\ 
& \textbf{G(1,0.005)} & 26471.83 & 2295e+3 & 1093e+2 & 1126e+7 & 3253.86 & 7046e+6 \\ 
& \textbf{G(1,1)} & 3916e+9 & 3698e+6 & 2868.80 & 1119e+5 & 1931e+4 & 4354e+6 \\ 
\midrule 
\textbf{NB}& --& 3210.01 & 3170.35 & 3165.34 & 3181.61 & 3165.12 & 3165.19 \\ 
\textbf{Poisson}&-- & 2997.78 & 2995.40 & 2995.13 & 2995.65 & 2995.19 & 2995.16 \\ 
\bottomrule 
\end{tabular} 
\end{scriptsize} 
\end{table} 
We can conclude some points according Tabels \ref{Tab3} and \ref{Tab4}. 
The GC model, with priors $PC(l1)$ for $\alpha$ and $Gamma(0.01,0.01)$ for $\tau$, demonstrates a better fit than others in terms of WAIC and LS criteria. Afterwards, GC model with priors $PC(l1)$ for $\alpha$ and $SD$ for $\tau$ has good fitting. In previous literature, this dataset called as an under dispersed data. Our obtained results confirm this; the GC model represents the best fit than the Poisson and the NB models. 
Moreover, we prepare posterior inference for model parameters under different models with these priors ($PC(l1)$ for $\alpha$ and $SD$ for $\tau$) in Table \ref{Tab6}. 
The estimate of the dispersion parameter $\alpha$ for the GC model illustrates an underdispersion in the data; however, the corresponding $95\%$ credible interval shows the Poisson model could also be the appropriate alternative for these data. 
The results of fitting show that smoking consumption has a significant even so slight positive effect on the mortality rate under three models. For the three proposed models, there is a negligible difference between the regression parameter estimates. Further, the estimates of the precision parameter, $\tau$, for the GC and the Poisson models are the same, with a slightly larger value for the NB model. 
\begin{table}[h!] 
\begin{footnotesize} 
\centering \caption{\label{Tab6} \footnotesize{Summary of posterior estimates, mean ($2.5\%$, $97.5\%$), for the fitted models of larynx data}} 
\begin{tabular}{cccccc} 
\toprule 
\textbf{Model} & $\alpha$ & $\tau_{\phi}$ & $\beta_0$ & $\beta_1$ & \textbf{ DIC} \\ 
\midrule 
\textbf{GC} & 1.882 & 0.673 & 1.777 & 0.008 & {\bf 2839.541} \\ 
& (0.936, 3.048) & (0.571,0.775) & (1.494, 2.058) & (0.002, 0.014) &\\ 
\cmidrule{2-6} 
\textbf{Poisson}& --- & 0.714 & 1.740 & 0.009 & 3068.31 \\ 
& & (0.606,0.826) & (1.447, 2.033) & (0.003, 0.014) &\\ 
\cmidrule{2-6} 
\textbf{NB} & --- & 0.841& 1.777 & 0.008 & 3216.954 \\ 
& & (0.688,0.974) & (1.481, 2.072) & (0.003, 0.014) &\\ 
\bottomrule 
\end{tabular} 
\end{footnotesize} 
\end{table} 
Figures \ref{fig1} and \ref{fig2} represent the observed versus predicted values and fitted values respectively for the hold-out regions. The predicted values are computed according to the mean of the predictive distributions. From figure \ref{fig1} this is appears that for our 10 hold-out areas the $95\%$ prediction intervals are quite broad producing a $100\%$ empirical coverage for all three models. However, the length of the prediction intervals for the GC model is generally shorter than the other couple models. The results of figure \ref{fig2} are the same as \ref{fig1}. 
\section{Summary and Discussion}\label{Sec7}
  The Poisson regression model is a common practice in many applications for modelling counts. However, count data usually have various levels of dispersion and, consequently, this 
the inherent property of counts should be included in the model in other to achieve more reliable results. In this paper, we 
applied a gamma-count (GC) model as a consequence of the renewal theory that relates nonexponential waiting 
times between events and the distribution of the counts. Our proposed model has excellent flexibility due to analyzing data with various dispersion, under- or over-dispersion and is a generalization of the Poisson model, under certain conditions, becomes the Poisson model. 
 
Although, model fitting and inference in a Bayesian GC structured additive (GCSA) regression model can be carried out using MCMC 
methods, but for the proposed model it comes with severe problems regarding convergence and 
computational time. 
The INLA method (Rue et al. (2009)) can be used as an efficient method for deriving estimation posterior inferences. 
INLA is a user friendly, one merit of the INLA 
the approach is that there is a package, called R-INLA, that can be used in the free software R, 
Furthermore, consequently, practitioners have the methodology at their disposal. 
Priors have the prominent role in the Bayesian inferences and as Simpson \textit{et al.} (2017) point out, 
Regard to Simpson \textit{et al.} (2017), we get PC prior to the dispersion parameter of the GCSA model. 
In Bayesian analysis of the structured 
additive regression models, prior elicitation is, in particular, an issue 
with choosing hyperpriors for the smoothing variances. As Klein \textit{et al.} (2016), proposed scale-dependent prior to the smoothing variances, we use scale-dependent prior to precision parameters. Subsequently, we consider some alternative priors for sensitivity analysis. 
The outcomes  of the simulation study and real examples showed that in most cases, $PC(1)$ prior for dispersion parameter was superior to the others. Besides, scale-dependent prior has a good performance for the smoothing parameter as the same as $HC(0.022)$ and $Gamma(0.01,0.01)$. 

\end{document}